\documentclass[a4paper,fleqn,usenatbib]{mnras}

\usepackage{float} 
\usepackage[T1]{fontenc}
\usepackage{ae,aecompl}

\usepackage{graphicx}	
\usepackage{amsmath}	
\usepackage{amssymb}	

\title[
Accretion disc modeling in UZ Ser \& CY Lyr 
]
{
The hydrogen Balmer lines and jump in absorption in accretion disc modeling 
- an ultraviolet-optical spectral analysis of the dwarf novae 
UZ Serpentis and CY Lyrae  
}

\author[P. Godon et al.]{
Patrick Godon,$^{1,2}$\thanks{E-mail: patrick.godon@villanova.edu }
Edward M. Sion,$^{1}$ Paula Szkody,$^{3}$ William P. Blair$^2$ 
\\
$^{1}$Department of Astrophysics and Planetary Science, Villanova University, 800 E. Lancaster Ave., Villanova PA 19085, USA \\ 
$^{2}$Department of Physics and Astronomy, Johns Hopkins University, 3700 San Martin Dr., Baltimore, MD 21218, USA \\ 
$^{3}$Department of Astronomy, University of Washington, 3910 15th Ave NE, 
Seattle, WA 98195, USA \\ 
}

\date{Accepted XXX. Received YYY; in original form ZZZ}

\pubyear{2019}

\begin{document}
\label{firstpage}
\pagerange{\pageref{firstpage}--\pageref{lastpage}}
\maketitle

\begin{abstract}

The spectra of disc-dominated cataclysmic variables (CVs)  
often deviate from the spectra of accretion disc models;   
in particular,  the Balmer jump and absorption lines
are found to be shallower in the observations than 
in the models. 
We carried out a combined ultraviolet-optical spectral analysis of 
two dwarf novae: UZ Ser in outburst, decline, and quiescence, 
and CY Lyr on the rise to outburst and in outburst. 
We fit the Balmer jump and absorption lines, the continuum flux
level and slope by adjusting the accretion rate, inclination, 
and disc outer radius. 
For both systems we find an accretion rate 
$\dot{M} \approx 8 \times 10^{-9}M_\odot$/yr in outburst,  
and  $\dot{M} \approx 2-3 \times 10^{-9}M_\odot$/yr 
for the rise and decline phases.  
The outer disc radius we derive is
smaller than expected ($R_{\rm disc} \approx 0.2a$, where $a$ is
the binary separation), except during late rise  
(for CY Lyr) where $R_{\rm disc}=0.3a$. 
UZ Ser also reveals a 60,000~K white dwarf.
These results show that during a dwarf nova cycle  
the radius of the disc is the largest just before the peak of the 
outburst, in qualitative agreement with the disc instability model
for dwarf nova outbursts.
We suspect that an additional emitting component (e.g. disc wind) 
is also at work to reduce the slope of the continuum and size of the 
Balmer jump and absorption lines. 
We stress that both the outer disc radius and disc wind need  
to be taken into account 
for more realistic disc modeling of CVs.   

\end{abstract}


\section{Introduction} 
Cataclysmic Variables (CVs) are evolved compact binaries in which a 
white dwarf (WD) star (the primary) accretes matter from a main (or post-main) 
sequence star (the secondary) filling its Roche lobe. As a consequence,
matter streams from the secondary through the first Lagrangian point 
into the Roche lobe of the primary. In weakly- and non- magnetic CVs the stream 
of matter eventually forms an accretion disc around the WD
\citep[see][for a review]{war95}.
   
Dwarf novae (DNe) are a subclass of non-magnetic CVs found mostly 
in a state of low accretion, `quiescence', interrupted periodically or 
sporadically by periods of intense accretion or `outbursts'.  
During an outburst, the mass accretion rate in DNe reaches  
$\dot{M} \sim 10^{-8}-10^{-9} M_{\odot}$yr$^{-1}$, and the 
accretion disc peaks in the ultraviolet (UV). 
During quiescence, the mass accretion rate drops to   
$\dot{M} \sim 10^{-11}-10^{-12} M_{\odot}$yr$^{-1}$ and the
accretion heated WD, with a temperature $T_{\rm eff} \sim 15,000-50,000$~K,
becomes the dominant source of UV light in the system.
   
The remaining non-magnetic disc CVs form the nova-like (NL) subclass,  
and are found predominantly in a state of high accretion. 
In NLs in high state and DNe in outburst,  
the accretion disc dominates both the UV and the optical bands.  
For this reason, disc-dominated DNe in outburst and NLs in high state 
are  ideal laboratories to study accretion discs \citep[e.g.][]{war95}.  

However, it was shown \citep{wad84,wad88,lon91} 
that the UV spectral energy distribution (SED) 
of many
disc-dominated CVs systematically disagrees with the standard disc
model \citep{pri81} SED.  
Standard disc model SEDs  
\citep[such as e.g. in][]{lon91,wad98,lin12}  
were too blue compared to the UV data, with 
the NLs exhibiting a larger discrepancy than the DNe 
\citep{lad91,lin07,lin08,lin09,lin10,god17}. 
Alternatively, one can state that  
the standard disc models that scaled to the
correct distance to these CV systems 
did not provide acceptable chi square ($\chi^2$) fits. 

It was suggested 
\citep{lon94,pue07,lin10}  to revise the disc temperature
profile in the innermost part of the disc to help flatten the slope
of the UV continuum.
Quantitatively, the theoretical SED of the standard disc model was predicted 
to behave as $F_{\lambda} \propto \lambda^{-2.33}$
\citep[][
who explicitly mentioned 
{\it when neglecting the disc edges}]{lyn69}.
A full disc atmosphere calculation \citep[e.g.][]{wad98}, however, 
gives an  UV SED slightly steeper, with 
$F_{\lambda} \propto \lambda^{-2.5} - \lambda^{-3.0}$. 
While the difference in results is likely due, in part, to the black 
body (BB) 
assumption vs. disc atmosphere calculations, it is also   
due to the effects of the inner boundary condition
creating a maximum in the disc temperature at $R=1.36 R_0$, where
$R_0$ is the inner disc radius.
In  \citet{god17} we showed how the location of the inner radius
disc, where the boundary condition is imposed,
affects the slope of the FUV continuum  
(and in the present work, we show how the size of the outer radius
of the disc affects the continuum slope in the NUV and optical).  
Compared to this, DNe have an average UV SED  
$F_{\lambda} \propto \lambda^{-2.25}$, while NLs are flatter
with  $F_{\lambda} \propto \lambda^{-1.90}$ for UX UMa systems,
$ \propto \lambda^{-1.70}$ for VY Scl systems, and
$ \propto \lambda^{-1.63}$ for SW Sex systems
\citep[see][]{god17}.

At the same time, standard disc model optical spectra 
exhibit strong Balmer jumps in absorption which are not observed
\citep{wad84,lad89a,kni98}. Instead, a small Balmer jump is seen in absorption 
in the early phase of the outburst (rise) of DNe, it decreases with time 
and, as the system reaches quiescence, it  
is replaced by a Balmer jump in emission together with the low
order of the Balmer series also in emission
\citep[e.g.][]{ver84,szk85,tho98}.  
As to the disc-dominated NLs, most of the time their spectra show either 
no Balmer jump at all, or a Balmer jump in emission;  
and when seen in absorption the amplitude of the jump is rather small  
\citep[e.g. RW Tri, UX UMa;][]{wil83}. 

It has been suggested 
\citep[][but see also \citet{kni97}]{mat15} 
that an accretion disc wind in high-state
non-magnetic nova-likes could significantly contribute to the overall UV and
optical spectrum. The disc wind adds Balmer continuum emission 
(a Balmer jump in emission) and flattens the slope of the entire spectrum, 
thereby improving the spectral fit both in the UV and optical. 
It has also been proposed \citep{nix19} that large magnetically controlled
zones 
\citep[MCZs, due to the presence of an inverse cascade of magnetic 
helicity in three-dimensional MHD turbulence; e.g.][]{sha11}  
could transfer mass and angular momentum outwards,
decreasing the temperature in the inner disc ($R \lesssim 3 R_*$) 
while increasing it in the outer disc ($R >> R_*$), thereby
flattening the slope of the spectral continuum in these systems. 
The time-scale involved in that process is so large that it is predicted
to occur in nova-likes remaining in high state for very long periods of time.
If that's the case, one would expect UX UMa systems (that never
go into a low state) to have the flattest slope of all, however,
their slope (-1.90) is steeper than VY Scl systems (-1.70)
and SW Sex systems (-1.63).

In a previous work \citep{god17}, 
%
we pointed out an additional factor that has to be taken into account 
when modeling accretion discs:       
the increase of the inner radius of the disc (due, e.g., to the presence 
of a geometrically thick boundary layer) 
can significantly reduce the inner disc peak temperature,
resulting in a UV spectrum with a shallower continuum slope 
in better agreement with the observations.  

In the present work we further model the disc to investigate the
effect of the outer radius of the disc spectrum.  
We show that, at high accretion rate,
the size of the disc dictates the amplitude of the Balmer jump in absorption. 
The outer disc temperature, which is the coldest temperature in the
disc, increases with
decreasing radius and can be set to reduce the size of 
the Balmer jump in absorption to fit the optical data of some
disc-dominated CV systems.
The effect of varying the outer disc size on the spectrum was 
theoretically 
investigated in a more general way by \citet{sha91} and \citet{ida10}; 
it was also  
invoked to explain the poor fit of the standard disc model spectrum
to UV and optical spectra of some short period DNe \citep{has83}.  
Here, we carry out a UV-Optical spectral analysis of the two DNe UZ Ser and 
CY Lyr paying particular attention to the size of the outer radius of the disc.

The two DNe UZ Ser and CY Lyr are introduced in the next section; 
the archival data is presented in section 3; our spectral modeling
method is explained in section 4; the results are presented and discussed
in section 4, followed by a summary and our conclusions in section 6.   

\begin{table*}
\centering
\caption{Adopted System Parameters}
\label{tab:sys_par}
\begin{tabular}{cclcl}
\hline
Parameter                    & UZ Ser        & Reference      & CY Lyr  & References  \\ 
\hline
  P             (days)       & 0.17589       & \citet{ech88,ech07}  & 0.1591  & \citet{tho98}  \\ 
 $M_{\rm wd}$  ($M_{\odot}$) & $0.9\pm 0.1$  & \citet{ech07}  & $\sim 0.8$    & averaged CV WD mass \citep{zor11}  \\  
 $M_2$         ($M_{\odot}$) & $0.4 \pm 0.1$ & \citet{ech07}  &         &                \\  
 $i$           (deg)         & $50\pm 10$    & \citet{ech07}  & 35      & this work      \\  
 $E(B-V)$                    & 0.24          & \citet{bru94,god17}  & 0.15 & \citet{szk85}  \\
  $a$          ($10^5$km)    & $\sim 10$     &                & $\sim$9  &                \\ 
  d (pc)                     & $\sim$300-400  & see text       & $490\pm 10$ & Gaia DR2 \citet{pru16,bro18}  \\ 
\hline
\end{tabular}
\label{para} 
\end{table*}

\section{The dwarf novae UZ Serpentis and CY Lyrae} 

{\bf The Z Cam Dwarf Nova UZ Serpentis.} 
UZ Ser is a Z Cam subtype of DN, namely it exhibits phases when 
the system remains in an intermediate, ``standstill'', state  
\citep{dyc87,dyc89,hon98}. 
From the AAVSO we find that the system can reach a maximum visual magnitude
$m_v=11.7$ and a minimum as low as $m_v=17$.
The system parameters are not well known, but an analysis of the masses
and inclination \citep{ech07} favors a mass ratio $0.3 < q < 0.6$,  
a white dwarf mass $M_{\rm wd} =0.9 \pm 0.1 M_{\odot}$, 
and a secondary mass $M_{2} = 0.4 \pm 0.1M_{\odot}$. 
As the system does not show eclipses of any kind, it must 
have a moderate inclination $i < 65-70^{\circ}$, 
and the analysis of \citet[][Fig.5 in their manuscript]{ech07} implies $i > 38^{\circ}$  
(consistent with the amplitude of the optical modulation \citep{ech88}). 
Therefore, following \citet{ech07}, we assume here  
$i = (50 \pm 10)^{\circ}$. 
With a period of 0.17589 days \citep[a little above 4hr][]{ech07}, 
the binary separation is of the order of $a \sim 10^{11}$~cm. 

The exact distance to the system is not known,
as its parallax has not been obtained by Gaia nor by
Hipparcos. 
Assuming a negligible color excess, 
\citet{her44} suggested that UZ Ser is located in front of a cloud  
itself located at a distance of 200-300 pc producing a regional 
$E(B-V)$ of 0.4 \citep{bak42}. 
Namely, \citet{her44} suggested $d < 200$~pc.  
That region of the sky, however, is very {\it patchy}  
(as can be verified with the Galactic
dust reddening and extinction map \citep{sch98,sch11}
\footnote{\url{https://irsa.ipac.caltech.edu/applications/DUST/}}), 
and for that reason  the distance estimate from \citet{her44}
has been deemed unreliable \citep{ech81}. 
In addition, the reddening of UZ Ser might not be negligible,
it is likely of the order of E(B-V)=0.25 (see next section) and 
could be as large as 0.30-0.35 \citep{ver87}. 

We use a more recent extinction map for individual objects
\citep{mor15}
\footnote{\url{http://svo2.cab.inta-csic.es/theory/exmap/index.php}}.  
In the direct vicinity of UZ Ser (at an angular distance of 8' from UZ Ser) 
there are three objects with a total extinction $A(V)\approx 3.5$
(corresponding to $E(B-V)\approx 1.13$ if we assume $R=3.1$) located at 
2.323~kpc, 2.47~kpc, and $\sim 3$~kpc. 
These are consistent with the presence of 
a second cloud located at a distance of 1-2~kpc adding an extra color 
excess of +0.5 mag to the first cloud as noted by \citep{bak42}.  
Eight more objects within a one degree radius 
have a reddening of  about 0.4-0.8 
with distances between 1.6 and 3 kpc. 
Within a radius of 2~deg all the objects have $E(B-V) \ge 0.32$. 
There are only a handful of objects at $d < 1200$~pc
(and all with $d > 300$~pc)  
and their color excess (as a function of the distance) seems 
to follow a linear  pattern.  This could confirm the presence of a  
first cloud having a color excess of $\sim 0.3$, and would  
imply that UZ Ser, with a color excess of $\sim 0.25$, is 
at the distant edge of (or just behind) 
the first cloud at a distance $d \sim 300$~pc. 
However, due to the paucity of data points for $d<1200$~pc
and the already mentioned patchiness of the sky in this region, 
UZ Ser could also be located further away. 
\citet{urb06} used the correlation between the maximum magnitude  
of a CV systems and its orbital period to derive a distance of
about 280~pc.
It is most likely that $ 200$~pc$ < d < 500$~pc,  
here we assume: $ d \sim 300-400$~pc. 

We list all the system parameters we adopted for UZ Ser in Table \ref{para}. 

In the UV, UZ Ser has only been observed with the {\it International Ultraviolet
Explorer (IUE)} \citep{ech81,ver84}. 
The continuum slope of the (dereddened) {\it IUE} spectra in outburst 
is rather steep, -3 on a log-log scale [see  \citet{god17}].  
This makes UZ Ser ``bluer'' than all CVs in outburst,
and indicates the presence of a hot component, possibly the disc at a high 
$\dot{M}$. Even its quiescent IUE spectrum (obtained on Aug 15, 1982) 
has a rather steep slope, likely
due to the heated WD. While most CVs have disc emission with a UV spectral
continuum slope too shallow compared to the standard disc model
\citep{pue07,god17}, UZ Ser
has a UV continuum slope comparable to a standard disc model accreting
at a high mass accretion rate. Also it exhibits a Balmer Jump of 0.3 mag
\citep{pan79} in absorption in the optical, in agreement with
the standard disc model and contrary to many disc-dominated CVs. 
UZ Ser seems to be one of a few systems exhibiting spectral
features in good agreement with the standard disc model.  

The ephemeris of the system 
is HJD = 2446622.68149(5) + 0.17589(2)E \citep[phase zero when the 
backside of the secondary is facing the observer;][]{ech07}. 
However, the error in the period (0.00002) and the number of 
cycles ($\sim 10,0000$) since the {\it IUE} observations 
(HJD $\sim 2444463-2445197$) gives a large orbital phase error
$\sim \pm 0.25 - \pm 0.16$, making an orbital phase
variability analysis impossible. The low S/N of the {\it IUE} spectra 
also prevents an  assessment of the absorption line radial velocity.

{\bf The U Gem Dwarf Nova CY Lyrae.}  
With a period of 0.1591 days \citep[a little less than 4hr;][]{tho98}, 
CY Lyr is classified as a U Gem type dwarf nova. It exhibits   
relatively short outbursts lasting about 5 days \citep{szk84},
during which its magnitude increases from 17.0 to 13.2 \citep{szk85}.   
Its inclination is unknown, but with a radial velocity $K_1=126 \pm 7$~km/s 
\citep{tho98} and showing no eclipse of any kind, it is likely
that the system is viewed at a moderate inclination. 
CY Lyr has a Gaia parallax of $2.04 \pm 0.04$ mas 
\citep{pru16,bro18,eye18,lur18}, giving a distance $d= 490 \pm 10$~pc.  
The secondary star was estimated to be an M3.5 dwarf assuming a distance
of 320~pc \citep{tho98}, but the larger Gaia distance  implies that
the M dwarf is of an earlier type. 
The mass of the WD is unknown and in the present work we will assume 
an average CV WD mass of $\approx 0.8 M_{\odot}$ \citep{zor11}. 
Under these assumptions, the binary separation of the system
is likely similar to that of UZ Ser, and because of its slightly
smaller binary orbital period, we assume here $a = 9 \times 10^5$km.   
All the systems parameters are listed in Table \ref{para}. 

CY Lyr has a reddening $E(B-V)=0.15$, estimated from the 2175~\AA\ ISM
absorption feature in its {\it IUE} spectrum \citep{szk85}.  
The continuum slope of its dereddened {\it IUE} spectrum in outburst 
is -2.5 \citep[on a log-log scale][]{god17}, in good agreement
with standard disc model spectra \citep{wad98}.

\clearpage

\section{The Archival Data}

\subsection{UZ Serpentis.} 
UZ Ser has a total of 16 {\it IUE} spectra (8 SWP \& 8 LWR) 
covering both its outburst and quiescent states. 
A first (SWP+LWR) spectrum was obtained on 1980 August 12, 
a day after the system went into a short outburst \citep{ech81}.
Another single (SWP+LWR) spectrum was obtained on 1981 September 22,
within a day of the peak of an outburst.  
A series of 6 (SWP+LWR) spectra were obtained 
during the second week of August 1982 as the system was in outburst
and declined into quiescence \citep{ver84}. 
Details on the archival {\it IUE} observation are given in Table \ref{iue_log}.    

The {\it IUE} Aug 1982 observations were coordinated with simultaneous 
spectroscopic optical observations \citep{ver84} and are therefore of special interest. 
On 1982 Aug 6 UZ Ser went into outburst
(reaching a magnitude of 12.8) and the {\it IUE} spectra were obtained 
on Aug 8, 9, 10, 11, 12 \& 15, as the system faded back into quiescence.   
The continuum flux level decreased by a factor of 20 from Aug 8 to Aug 15. 
We digitally extracted the optical spectra obtained on Aug 11 \& 15
from \citet{ver84},
as they coincide (to within about 1/2~hr) with the {\it IUE} spectra
obtained on these dates (see Table \ref{opt_timing}). 
Even though \citet{ver84} did collect more optical data, we were only
able to extract those spectra that were displayed in the original publication.   

Of all the {\it IUE} exposures, the 22 Sep 1981 spectrum (SWP15078 \& LWR11605) 
has the highest continuum flux level, about 25 times larger than the 
15 Aug 1982 spectrum (SWP17700 \& LWR13960) which has the lowest continuum flux level.
The AAVSO data indicate that the 
system possibly reached a visual magnitude of 12.8 on Sep 22, 1981. 
There is no valid AAVSO data for 15 Aug 1982, but the AAVSO data show a decline
from Aug 9 ($m_v=13$) through Aug 12 ($m_v=15$), and the {\it IUE}
data show that the UV continuum
flux level decreased  from $4.5 \times 10^{-14}$erg/s/cm$^2$/\AA\  
on Aug 12 and reached a minimum of $1.5 \times 10^{-14}$erg/s/cm$^2$/\AA\  
on Aug 15 (a drop by a factor of 3). On that day, Aug 12, the visual magnitude
was probably close to 17.

Therefore, it is reasonable to assume that the Sep 1981 
spectrum is mostly dominated by emission from the accretion disc 
(mid-outburst), while the Aug 15 spectrum is likely
dominated by emission coming mostly from the WD (quiescence).

The mean extinction laws of \citet{sav79} and \citet{sea79} 
are commonly used to deredden spectra affected by interstellar
extinction, or equivalently the formulation of \citet{car89a,car89b}
is used (with a value of $R_v$ near 3.1-3.2 \citep{mat90}).
In our more recent work we dereddened UV and optical spectra using 
the extinction law given by \citet{fit07}.

The interstellar extinction produces a strong and broad absorption
feature centred at 2175~\AA , due mainly to polycyclic aromatic
hydrocarbon (PAH) grains \citep{li01}.
PAHs, however, do not dominate the FUV extinction, and the 2175~\AA\  
bump correlates poorly with the FUV extinction \citep{gre83}:                   
there is a large sample variance about the mean average Galactic extinction
curve observed in the shorter wavelength of the FUV.   
Nevertheless, the reddening is often assessed from the 2175~\AA\ PAHs feature 
and gives an accuracy of about 20\%. 
Furthermore, the extinction curve itself is an average throughout the 
Galaxy and could be different in different directions.

All the above extinction laws have a similar correction from the 2175~\AA\ bump
longward of 1,500~\AA , but at shorter wavelength 
Cardelli's gives a higher correction, Seaton and Savage \& Mathis 
gives a smaller correction, and Fitzpatrick \& Massa's gives an 
intermediate correction \citep[as already noted by ][]{sel13}. 
This discrepancy increases with decreasing wavelength.
It has been shown \citep{sas02} that in the FUV 
the observed extinction curve is actually consistent with an extrapolation of the
standard extinction curve of \citet{sav79}. 
Consequently, in the present work, we slightly modify our dereddening
software (based on \citet{fit07} analytical expression) to agree with
an extrapolation of \citet{sav79} in the FUV range.

Based on the 2175~\AA\ absorption feature, 
the reddening toward UZ Ser has been estimated to be as low as 0.1
\citep{ech81} and as large as 0.35 \citep{lad91}.  
We derive a reddening value E(B-V)=0.24 for the Sep 1981 
{\it IUE} spectrum (with the highest continuum flux level), 
and E(B-V)=0.30 for the Aug 08, 1982 spectrum (when the system was 
in outburst). The region of the LWR {\it IUE} spectra in the vicinity
of the 2175~\AA\ absorption feature is rather noisy with what
looks like deep absorption lines. While some of these absorption features  
might be actual absorption lines (e.g. Si\,{\sc i} 1983-6, 
He\,{\sc ii} 2053 \& 2306,
C\,{\sc iii} 2297), other features might be artifacts. 
Binning the spectra to 15~\AA\ to derive the reddening 
\citep[a usual practice;][]{ver87} might therefore over estimate
the reddening by increasing the 2175~\AA\ feature with these absorption
lines and artifacts.  We here adopt the E(B-V)=0.24 value as it
was obtained from the most reliable spectrum (highest continuum flux level)
and it is also consistent with the (average) value given in \citet{bru94}.    
We note that \citet{ver84} also found E(B-V)=0.30 for the Aug 08, 1982 
spectrum. The LWR spectra obtained in late decline and quiescence are
far too noisy to derive E(B-V) from the 2175~\AA\ absorption feature.  
We deredden the spectra assuming E(B-V)=0.24.

\begin{figure}
\vspace{1.0cm} 
\includegraphics[width=\columnwidth]{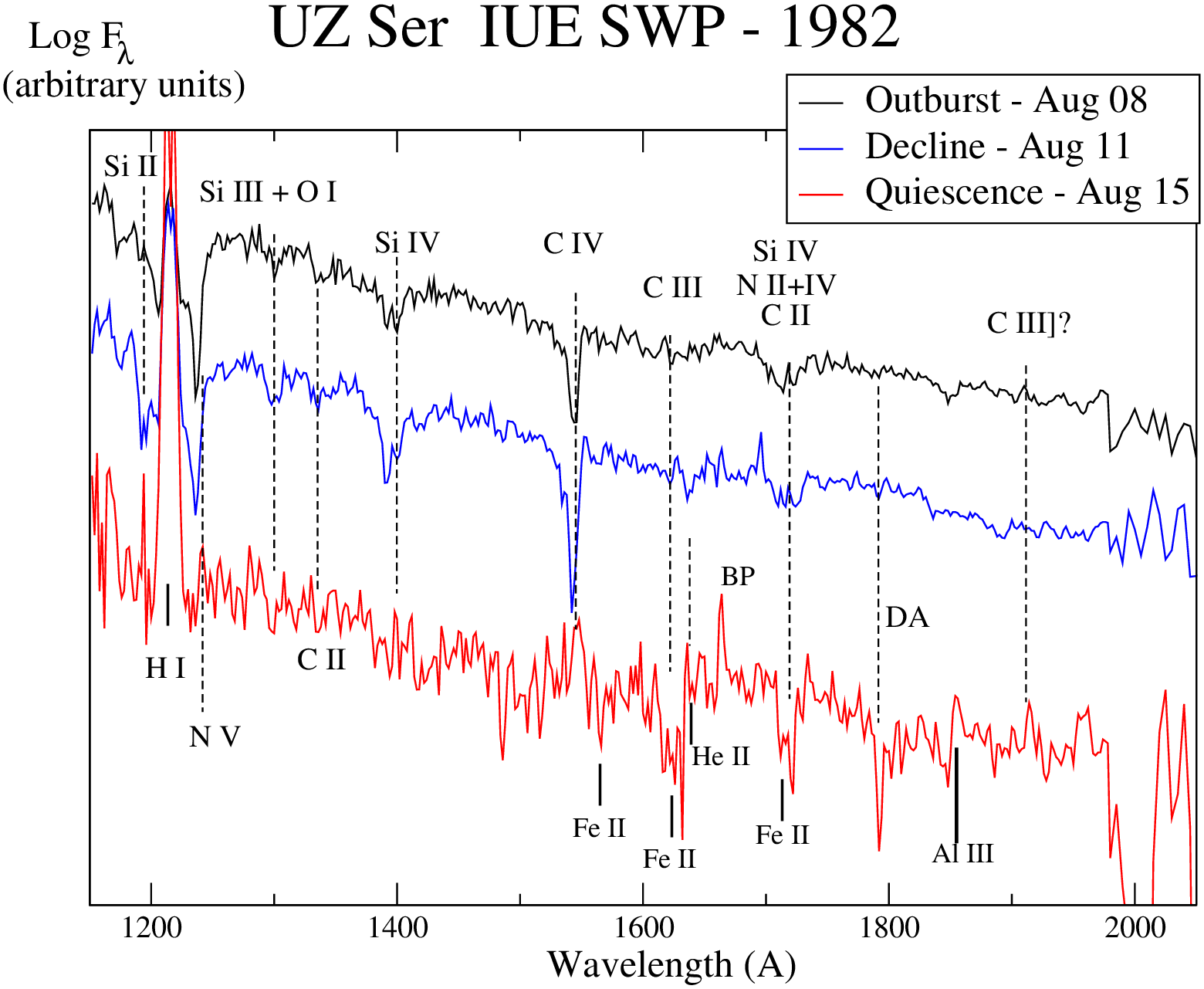} 
\vspace{-4.5cm} 
\caption{
The dereddened 
{\it IUE} SWP Spectra of UZ Ser in outburst (SWP17633/Aug 08, in black), decline
(SWP17661/Aug 11, in blue) and  quiescent (SWP17700/Aug 15, in red)  
are shown (for clarity) on an arbitrary logarithmic scale 
($Log(F_{\lambda})$, where $F_{\lambda}$ is in erg/s/cm$^2$/\AA ).  
The most common absorption lines have been 
marked. There is also a possible forbidden C\,{\sc iii} 
emission line around $\sim$1910~\AA .  
The quiescent spectrum exhibits some absorption features that
appear to be consistent with absorption from an iron curtain
(marked with Fe\,{\sc ii} below the spectrum). 
There is a detector artifact (``DA'') just before 1800~\AA ,
and a bad pixel (``BP'') near 1660~\AA .  
\label{u_id} 
}
\includegraphics[width=\columnwidth]{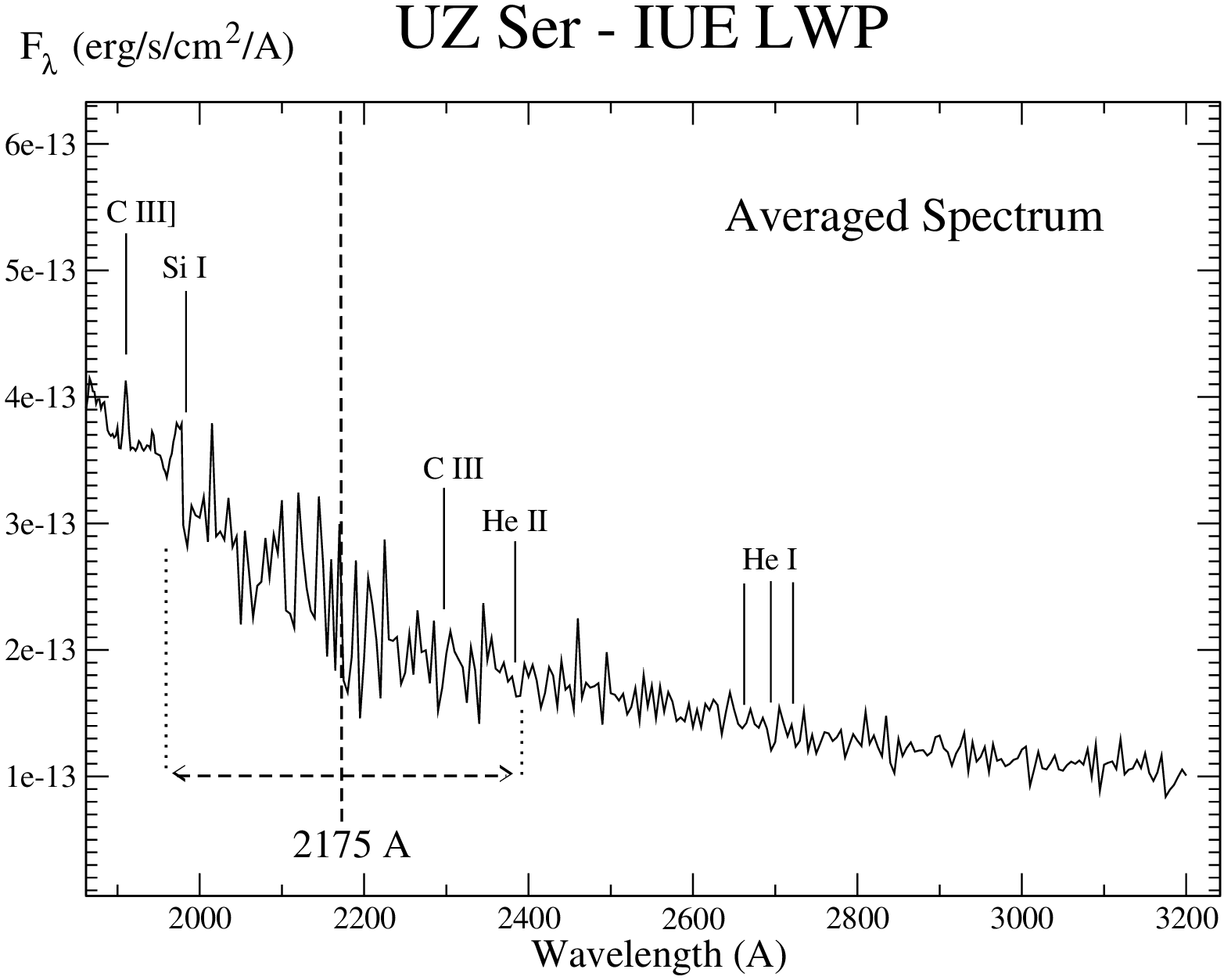}                   
\vspace{-3.5cm} 
\caption{
The dereddened 
{\it IUE} LWR co-added spectrum of UZ Ser. The 8 LWR spectra have been
co-added to increase the S/N.  We tentatively identify some absorption lines
as marked. The spectra were dereddened using the ISM polycyclic
aromatic hydrocarbon (PAH) 2175~\AA\ 
absorption feature. The location of the 2175~\AA\ feature is 
marked with a vertical dashed line. Note that the spectrum is very noisy 
within $\pm 200$~\AA\ in this region (marked with arrows), 
making the dereddening more difficult.  
\label{u_l_id} 
}
\end{figure}

We present the SWP15078 \& SWP17700 spectra in Fig.\ref{u_id} identifying
some absorption and emission lines. 
The outburst spectrum displays absorption lines from
Si\,{\sc iii} (1192~\AA),  
N\,{\sc v} (1240~\AA), 
C\,{\sc ii} (1335~\AA),  
Si\,{\sc iv} ($\sim$1400~\AA), 
C\,{\sc iv} ($\sim$1550~\AA), 
and  C\,{\sc iii} (1621~\AA); 
the absorption feature around 1303~\AA\ is a combination of  
Si\,{\sc iii} and  O\,{\sc i} lines.  The depression in the continuum
flux near 1720~\AA\ is due to  
N\,{\sc iv} (1719~\AA), but also likely to be  
Si\,{\sc iv} (1723~\AA),  
C\,{\sc ii} (1721~\AA ), and Si\,{\sc ii} (1711~\AA ).   
There is also a possible absorption line of O\,{\sc v} (1371~\AA ) as well 
as a forbidden emission line of C\,{\sc iii}] (1909~\AA).

\begin{table*} 
\centering
\caption{Archival {\it IUE} Observation Log}
\label{tab:obs_log}
\begin{tabular}{cccccccc}
\hline
System &  Short   & Long   & Date (UT)  & Time (UT)  & Exp. Time & SWP Flux   & Fig.      \\ 
Name &  Wave     & Wave   & mmm-dd-yyyy & hh:mm:ss   & seconds   & (relative) & \#       \\             
\hline
UZ Ser & SWP09769  & LWR08489 & Aug 12 1980 & 00:31:42 & 2400/1200   & 15     &                               \\ 
       & SWP15078  & LWR11605 & Sep 22 1981 & 18:41:43 & 2400/2100   & 23     & \ref{u_d1.0},\ref{u_d1.2}    \\ 
       & SWP17633  & LWR13901 & Aug 08 1982 & 19:02:28 & 2400/3600   & 20     & \ref{u_id},\ref{u_f_all}     \\ 
       & SWP17645  & LWR13909 & Aug 09 1982 & 23:43:46 & 3600/3300   & 17     &                                 \\ 
       & SWP17652  & LWR13917 & Aug 10 1982 & 19:13:49 & 4200/4200   & 14     &                                 \\ 
       & SWP17661  & LWR13922 & Aug 11 1982 & 18:52:20 & 4800/3600   & 7.4    & \ref{u_id},\ref{u_f_all}     \\ 
       & SWP17672  & LWR13930 & Aug 12 1982 & 20:22:17 & 7800/4800   & 3.3    &                                 \\ 
       & SWP17700  & LWR13960 & Aug 15 1982 & 18:20:10 & 10800/5400  & 1.0    & \ref{u_id},\ref{u_f_all}     \\ 
CY Lyr & SWP21030  & LWR16779 & Sep 12 1983 & 22:09:45 &  3600/2400  & ---    & \ref{c_id},\ref{c_all},\ref{c_f_all}    \\ 
       & SWP21058  & LWR16794 & Sep 15 1983 & 03:10:16 &  2520/1800  & ---    & \ref{c_id}                      \\ 
\hline
\end{tabular}
\label{iue_log} 
\end{table*}

\begin{table*} 
\centering
\caption{Optical Data Timing}
\label{tab:obs_tim}
\begin{tabular}{rccccccc}
\hline
System &  Date (UT) & RJD$^a$  & RJD$^a$ & RJD$^a$ & Optical Flux$^b$ & Fig.    \\ 
Name & mmm-dd-yyyy & {\it IUE} SWP & {\it IUE} LWR & Optical & (relative) & \#   \\             
\hline
UZ Ser & Aug 08 1982 & 45190.32 & 45190.37 &    ---   & --- &                             \\ 
       & Aug 09 1982 & 45191.53 & 45191.47 &    ---   & --- &                              \\ 
       & Aug 10 1982 & 45192.40 & 45192.35 &    ---   & --- &                              \\ 
       & Aug 11 1982 & 45193.39 & 45193.33 & 45193.36 & 5.0 & \ref{u_f_all}              \\ 
       & Aug 12 1982 & 45194.44 & 45194.50 &    ---   & --- &                              \\ 
       & Aug 13 1982 &  ---     &   ---    & 45195.41 & 2.0 &                              \\  
       & Aug 15 1982 & 45197.39 & 45197.46 & 45197.37 & 1.0 & \ref{u_f_all}              \\ 
CY Lyr & Oct 09 1982 & ---      &  ---     & 45252.00 & 1.05 & \ref{c_all},\ref{c_f_all}  \\ 
       & Jun 25 1997 & ---      &  ---     & 50625.00 &     &                              \\ 
       & Jun 29 1997 & ---      &  ---     & 50629.00 &     &                              \\ 
       & Jun 30 1997 & ---      &  ---     & 50629.67 &     &                              \\ 
       & Jun 30 1997 & ---      &  ---     & 50629.74 &     &                              \\ 
       & Jun 30 1997 & ---      &  ---     & 50629.79 &     &                              \\ 
       & Jun 30 1997 & ---      &  ---     & 50629.85 &     &                              \\ 
       & Jun 30 1997 & ---      &  ---     & 50629.90 & 0.41 & \ref{c_all}                 \\ 
       & Jun 30 1997 & ---      &  ---     & 50629.97 & 0.66 & \ref{c_all},\ref{c_f_i}   \\ 
       & Jul 01 1997 & ---      &  ---     & 50630.85 & 1.0 & \ref{c_all},\ref{c_f_all}  \\ 
\hline   
\end{tabular} \\ 
(a) RJD=relative Julian Date starting at JD=2,400,000.  
(b) The relative optical flux is measured in the vicinity $\lambda \in [4000,5000]$\AA\  
and it is different for each system.  
\label{opt_timing} 
\end{table*}

The quiescent spectrum has a lower S/N and
the identification of lines is less reliable.
It shows many lines in emission and fewer lines in absorption. 
We identify the Si\,{\sc iii} + O\,{\sc i} ($\sim 1300$~\AA) and 
C\,{\sc ii} (1335~\AA) absorption features, as well as 
the following emission lines: 
Ly\,{\sc $\alpha$} (most probably geo-coronal in origin), 
N\,{\sc v} (1240~\AA), Si\,{\sc iv} (1400~\AA), 
C\,{\sc iv} (1550~\AA ), He\,{\sc ii} (1640~\AA), 
Al\,{\sc iii} (1855~\AA), and the forbidden C\,{\sc iii}] (1909~\AA).   
The spectrum presents several regions with broad absorption features,
which could be due to C\,{\sc iii} (near 1620~\AA)
and N\,{\sc iv} (near 1720~\AA).  
However, these absorption features are very broad, and are more likely due to 
an absorbing ``iron curtain'' \citep{hor94}.  We identify three such regions 
where there is a multitude of Fe\,{\sc ii} lines: near $\sim$1560-1590~\AA , 
$\sim$1612-1635~\AA ,  and $\sim$1709-1725~\AA. The iron curtain is known
to dominate the region between $\sim$1500~\AA\ and $\sim$ 1700~\AA\ 
\citep[e.g.][]{smi94,pal17}. 
The last two regions could also have absorption lines from 
C\,{\sc iii}, Si\,{\sc iv} and N\,{\sc iv} as in the outburst spectrum.  

Veiling curtains are fairly common in spectra of CVs and could be due
e.g. to stream disc overflow material \citep{god19} and/or to an
outflow \citep[disc wind;][]{kni97}. 
The fact that we derive a reddening of 0.24 for the Sep 1981 spectrum
and 0.30 for the Aug 08, 1982 spectrum is a possible indication that the 
source is subject to intrinsic reddening. This would also be consistent
with veiling.  
As a consequence, in the Results Section we also discuss how a possible
uncertainty of about $\pm 0.05$ in the reddening, say $E(B-V)=0.2$ vs.
$E(B-V)=0.3$, translates into an uncertainty in the parameters
derived from our spectral analysis.

\begin{figure}
\vspace{0.5cm} 
\includegraphics[width=\columnwidth]{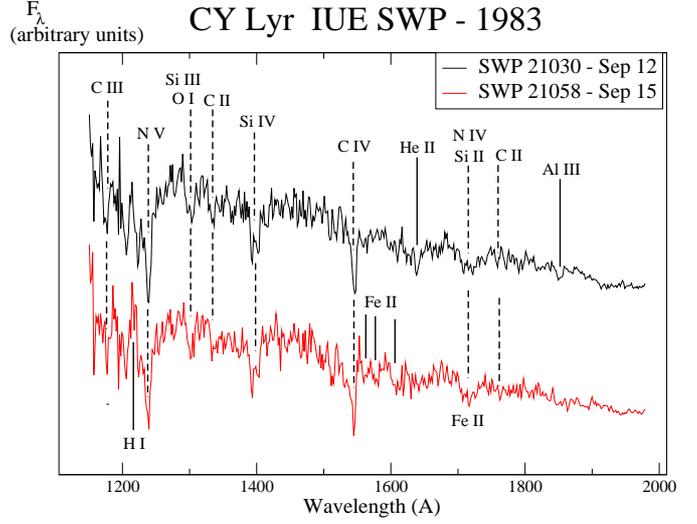}             
\vspace{-4.5cm} 
\caption{
The dereddened {\it IUE} SWP spectra ($F_{\lambda}$ vs. $\lambda$) 
of CY Lyr in outburst displayed on an arbitrary scale.
The Sep 15 spectrum is in red and the Sep 12 spectrum is in black.
Though the two spectra have about the same flux level, for clarity,
they have been displayed on an arbitrary scale so as to not
overlap. The most common absorption lines have been identified
as marked.  The Sep 15 spectrum has a continuum flux level about
10-15\% lower than the Sep 12 spectrum and displays some 
additional absorption lines reminiscent of the iron curtain
(see Fig.\ref{u_id} for comparison).  
\label{c_id} 
}
\end{figure} 

\begin{figure}
\vspace{-3.5cm} 
\includegraphics[width=\columnwidth]{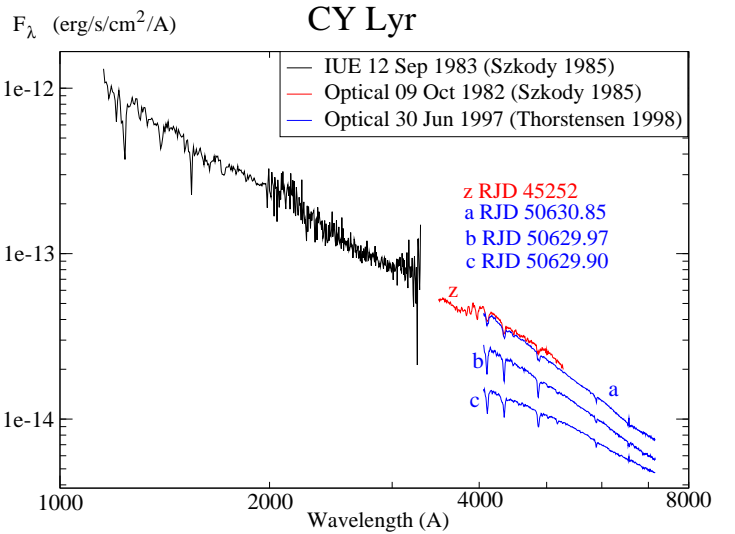}                     
\caption{
The optical spectra of CY Lyr compared to the {\it IUE} spectrum.
All the spectra have been dereddened.  
The {\it IUE} spectrum was obtained near outburst maximum light, the
1982 optical spectrum (z) was obtained during outburst, and the
1997 optical spectra were obtained on the rise to outburst 
(b \& c) and in outburst (a).  
\label{c_all} 
}
\end{figure}

\clearpage

\subsection{CY Lyrae.} 
CY Lyr was observed in the FUV with {\it IUE} 4 days into outburst
(Sep.12-15, 1983, see Table \ref{iue_log}),  as well as in the optical ($\sim 3,500-5,300$~\AA ) 
both in quiescence (Oct. 6, 1982) and in outburst \citep[Oct. 9, 1982][see Table \ref{opt_timing}]{szk85}. 
The optical spectra during the rise to outburst and decline
show narrow emission-line cores within the broad Balmer absorption. 
On Sep. 15 the {\it IUE} spectrum reveals a P Cygni feature in C\,{\sc iv} (1500~\AA) 
with the absorption component stronger than the emission component 
(with no emission component from other lines) 
which is not seen on Sep 12 at the beginning of the outburst
(see Fig.\ref{c_id}). The {\it IUE} LWR spectra of CY Lyr are
{\it unremarkable and rather noisy}, but provide a continuum flux
level that can be used in the spectral modeling.  
The outburst {\it IUE} and optical spectra are presented together 
in Fig.\ref{c_all}.  

CY Lyr was also observed in the optical by \citet[][on 7 \& 8 Aug 1981]{ech83} 
who presented a spectrum that just covers the Balmer Jump from
$\sim 3,800$~\AA\ to $\sim 5,500$~\AA\ with a continuum flux level
about twice as large as in \citet{szk85}, but with a lower resolution. 
The absorption lines depth and Balmer Jump amplitude (also in absorption)
are similar in both observations, however, the optical spectrum 
in \citet{szk85} covers the continuum down to $\sim 3,500$~\AA\ , 
which allows a better assessment of the amplitude of the Balmer Jump as
it includes the continuum between 3,500~\AA\ and 3,800~\AA\ that can be
modeled.  

CY Lyr was further observed at the end of June 1997 by \citet[][see Table \ref{opt_timing}]{tho98},  
who obtained 8 optical spectra ($\sim 4,000-7,300$~\AA ) almost every hour while
the system evolved from quiescence into outburst. The quiescent state 
spectrum of CY Lyr is typical of quiescent dwarf novae with sharp emission lines, and
the secondary is believed to contribute a significant fraction of the 
continuum flux level ($\sim 1/6$ at 6,500~\AA ). 
Besides a component from the secondary, the quiescent optical spectrum also
contains emission lines and possibly some additional continuum contribution.  
As the system rises into outburst, the continuum flux level increases, 
the spectrum becomes bluer, and the emission lines are replaced
by absorption lines. The absorption lines are strongest before 
maximum light, though H$\alpha$ does display very weak emission.
In Fig.\ref{c_all} we display the last 3 optical spectra as the system
reached its outburst level.  Spectrum (a) is one day into outburst and
matches remarkably well the optical spectrum (z) [also shown here in 
Fig.\ref{c_all}] from \citet{szk85}.  
Spectrum (b) was obtained $\sim$21hr before (a), and spectrum (c) was obtained 
$\sim$1hr40m before (b). Spectrum (b) displays the deepest absorption lines
and is likely to have no or minimum emission lines.  

These observations reveal that  
during quiescence the Balmer lines and jump are seen in emission, 
and the spectral slope of the continuum is rather flat.
As the system rises into outburst, the spectral slope becomes steeper
(the spectrum becomes bluer), 
and absorption lines start to dominate the spectrum. 
The absorption lines are strongest before the system reaches its 
maximum, then as the system remains in outburst, narrow emission line  
cores appear within the broad Balmer absorption and the C\,{\sc iv}
line starts to display a P Cygni profile. 
These are signs of an outflow (a possible disc wind), and, together 
with the absorption features from a possible ``iron curtain'' (Fig.\ref{c_id}), 
we suspect that here too (as for UZ Ser) there is material over the
disc in the line of sight of the observer originating from either a disc wind 
and/or stream disc overflow, or both.

\clearpage

\section{Spectral Modeling} 

\subsection{TLUSTY}
              
We use the FORTRAN suite of codes TLUSTY, SYNSPEC, ROTIN and 
DISKSYN \citep[often simply referred to as TLUSTY][]{hub88,hub94,hub95}  
to generate synthetic spectra of stellar atmospheres and discs.
The codes include the treatment
of hydrogen quasi-molecular satellite lines (low temperature),
and LTE/NLTE options. 
TLUSTY generates a stellar atmosphere model for a given 
effective surface temperature $T_{\rm eff}$ (in K), gravity $Log(g)$
(Log of cgs), and chemical composition (in solar units).   
Using the output from TLUSTY, SYNSPEC generates a continuum
with absorption lines. In the present case we do not generate emission lines. 
For disc spectra we use solar abundances 
and for stellar spectra we vary abundances
as needed. 
An introductory guide, a reference manual and an operational manual
to TLUSTY and SYNSPEC have been released and are available for 
full details of the codes \citep{hub17a,hub17b,hub17c}.

\subsection{Synthetic WD stellar atmosphere spectra} 
              
The code TLUSTY is first run to generate a one-dimensional (vertical)
stellar atmosphere structure for a given surface gravity ($Log(g)$),
effective surface temperature ($T_{\rm eff}$) and surface composition 
of the star. 
Computing a single model is an iterative process that is brought to convergence.
In the present case, we treat hydrogen and helium explicitly, 
and and treat nitrogen, carbon and oxygen implicitly 
\citep{hub95}. 
                
The code SYNSPEC is then run, 
using the output stellar atmosphere model from TLUSTY
as an input, and generates a synthetic stellar spectrum over a given
(input) wavelength range. Here we use its full capability and set
the spectral range to cover the FUV and optical: 
from 900~\AA\  to 7500~\AA . 
The code SYNSPEC derives the detailed radiation and flux distribution of 
continuum and lines and generates the output spectrum \citep{hub95}.           
SYNSPEC has its own chemical abundances input to generate lines for 
the chosen species.  
For temperatures above 35,000~K, we turn on
the approximate NLTE treatment of lines 
in SYNSPEC. 
   
Rotational and instrumental broadening as well as limb darkening 
are then reproduced using the routine ROTIN. 
Synspec also generate the specific intensities for a given angle
of inclination, which can then be translated into flux.  
In this manner we have already generated a grid of WD synthetic
spectra covering a wide range of temperatures and gravities 
with solar composition. 

For temperature below $\sim$10,000~K (and low gravity, $Log(g) \sim 4$), 
we turn on the Grey stellar atmosphere option in TLUSTY/SYNSPEC. 
For temperatures below 6000~K,  
we use Kurucz solar abundances stellar atmosphere models \citep[e.g.][]{kur79}. 
We also use these low temperature solar-abundances stellar spectra 
to model the secondary and/or the outer accretion disc when the
temperature and the effective surface gravity
are much lower than that of the WD.  

\begin{figure} 
\vspace{-5.cm} 
\includegraphics[width=\columnwidth]{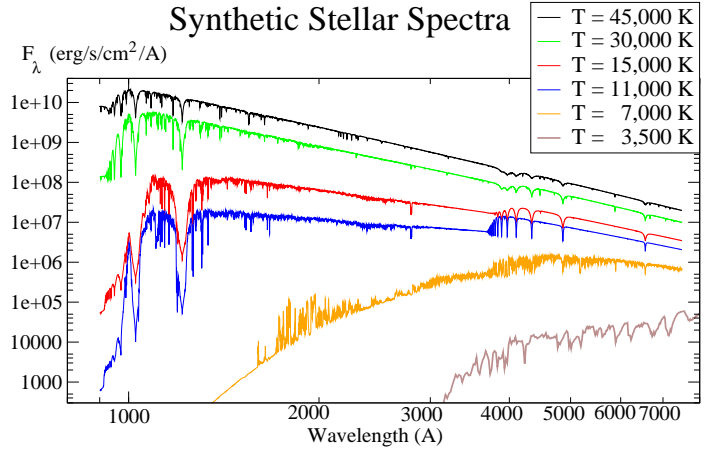}  
\vspace{0.5cm} 
\caption{
Stellar spectra ($F_{\lambda}$ vs $\lambda$ on a log-log scale) 
from 3,500~K to 45,000~K, for temperatures as shown in the upper right panel. 
The Balmer Jump in absorption, around  $\sim$3,500-4,000~\AA , 
is more prominent for temperatures in the range 10-15,000~K.  
We generate synthetic stellar spectra using TLUSTY/SYNSPEC from 
900~\AA\ to 7500~\AA\  for a temperature ranging from 
$\sim $10,000~K to $\sim$100,000~K, below
$\sim$10,000~K we use the Grey atmosphere option in TLUSTY/SYNSPEC,
below 6000~K we use  Kurucz models.  
\label{stellar_spectra} 
}
\end{figure}

In Fig.\ref{stellar_spectra} we present some synthetic stellar atmosphere 
spectra  for a temperature ranging from 3,500~K (with $Log(g)=4$) to
45,000~K (with $Log(g)=8$). 
The hydrogen Balmer absorption lines (series) form as the hydrogen atom 
electron transitions
from levels $n \geq 3$ down to level $n=2$. As $n$ approaches infinity, 
it creates a continuous absorption feature appearing as a jump (step) in the
continuum flux level in the vicinity $\lambda \approx 3750$ \AA\   
(Fig.\ref{stellar_spectra}). 
The Balmer jump in absorption
is especially prominent in stellar spectra 
when the temperature is in the range $8-15,000$~K. 
This is important, as the disc can be modeled as a collection of rings, 
each with a given temperature, and the resulting
amplitude of the Balmer jump  in the disc itself depends on the
lowest temperature in the disc (see next subsection). 

\subsection{Synthetic disc model spectra} 

Our disc model is based on the Shakura-Sunyaev
alpha disc model \citep[the {\it standard disc model,}][]{sha73,lyn74}.  
The disc is assumed to be optically thick, 
geometrically thin (in the vertical dimension),
rotating at a nearly Keplerian speed ($\Omega \approx \Omega_K$),   
and axi-symmetric. 
The energy dissipated by the Keplerian shear (between adjacent rings of
matter)  is radiated locally, and    
the resulting effective surface temperature 
is consequently solely a function of the radius $R$: $T(R)$  
\citep[one-dimensional, e.g.][]{pri81}.  

Since the disc has to adjust to the slowly rotating white dwarf,
its angular velocity has to decrease and, therefore,
a no-shear condition ($d \Omega/dR=0$) is set  
at the disc's inner boundary $R_0$. 
As a consequence, 
the temperature profile reaches a maximum at $r \approx 1.36 R_0$ 
and decreases to zero as $R \rightarrow R_0$. 
In the standard disc model: $R_0=R_{\rm wd}$.  
The validity of such an inner boundary condition might be questioned, 
especially since it dictates the angular momentum transfer (AMT)
from the disc to the star.  
A decreases in AMT, e.g. setting $R_0 \rightarrow 0$ \citep{lad89b} or  
$d\Omega/dR < 0$ \citep{mum19}, increases the luminosity and temperature
in the inner disc. In CVs the largest effect of such a boundary
condition is in the unobservable EUV wavelengths \citep{lad89b}. 
On the other hand an increase in AMT 
\citep[if one sets $R_0 > R_{\rm wd}$,][]{god17} will decrease 
the luminosity and temperature in the inner disc. 
The resulting slope of the disc spectral continuum is shallower for an
increased AMT and steeper for a decreased AMT. 
Here we simply assume the no-shear boundary condition
at the inner radius of the disc which is set at $R_0=R_{\rm wd}$. 

The standard disc model further assumes that the boundary layer
between the star and disc (where the remaining accretion energy 
is released) is very thin and radiates mainly in the X-ray
and EUV. In some previous work \citep[e.g. ][]{god12} we modeled 
the boundary layer as a hot inner ring in the disc, or alternatively
as a hot equatorial belt on the WD surface. This modeling is needed
and possible when analyzing {\it Far Ultraviolet Spectroscopic Explorer
(FUSE)} spectra of CVs (going down to 900~\AA ). However, the modeling diverges 
for {\it IUE} spectra, since the higher orders of the Lyman series 
(i.e. Ly$\beta$, Ly$\gamma$, ..) are needed to differentiate between
a hot WD, a fast rotating belt,  and/or a  hot ring/inner disc.  
In the present work, an increase of the WD temperature is enough
to help fit a steep spectral slope (see Results Section).   

In our modeling, 
the white dwarf mass ($M_{\rm wd}$), the mass accretion rate ($\dot{M}$),
the binary inclination ($i$), the inner disc radius ($R_0$), 
and the outer disc radius ($R_{\rm disc}$) are all input parameters.
Due to the tidal interaction of the secondary star, the maximum 
size of the accretion disc is expected to be between $0.3a$ 
(where $a$ is the binary separation)
for a mass ratio $q=M_2/M_1 \approx 1$, and about $0.6a$
for $q << 0.1$ \citep{pac77,goo93}, though in the present
work, $R_{\rm disc}$ is found by fitting the data.   

\begin{figure}
\vspace{-3.5cm}
\includegraphics[width=\columnwidth]{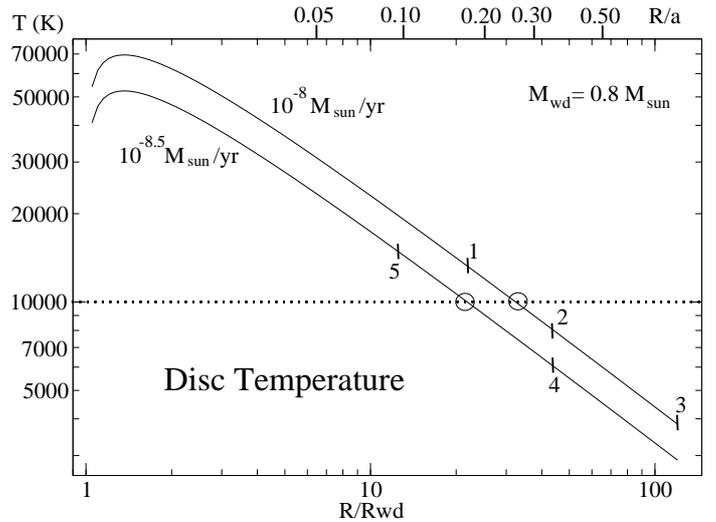} 
\vspace{-1.cm}
\caption{
The effective surface temperature of the accretion disc model  
is shown as a function of the radius on a log-log scale.  
Two mass accretion rates are displayed: 
$\dot{M}=10^{-8}M_{\odot}$/yr (upper graph), and  
$\dot{M}=10^{-8.5}M_{\odot}$/yr (lower graph).  
The central WD has a mass $M_{\rm wd}= 0.8M_{\odot}$ and a radius 
$R_{\rm wd}=7,000$~km. The location of the outer disc radius 
(e.g. 1, 2, 3, etc..) sets the lowest temperature in the disc. 
The horizontal dotted line shows the outer radii for T=10,000~K 
where it intersects the graphs. This is the temperature for which
the Balmer jump is the strongest. 
All the Wade \& Hubeny disc models (with spectral range 900 \AA - 2000 \AA )
are cut off at this radius since
the outer colder disc is not expected to contribute much 
at $\lambda < 2000$~\AA . 
A disc with an outer radius at
(e.g.) (1) or (5) will have a much smaller Balmer jump than a disc with 
an outer radius at (e.g.) (2), (3), or (4).  
Above the panel we indicate the size of the accretion disc as a fraction
of the binary separation ($a$) in UZ Ser.  
\label{disc_temp}
}
\end{figure}

In Fig.\ref{disc_temp} we display the temperature profile $T(R)$ for 
two accretion discs around a $0.8M_{\odot}$ WD with a radius 
$R_{\rm wd}=7,000$~km: 
one with $\dot{M}=10^{-8}M_{\odot}$/yr, and  
one with $\dot{M}=10^{-8.5}M_{\odot}$/yr.  
In these models the inner boundary of the disc is 
set at $R_0=R_{\rm wd}$ and the outer boundary can be set
at different radii (e.g. points 1, 2, 3, etc..). 
Setting the outer radius of the disc determines the lowest
temperature in the disc (note that because our disc rings have a 
finite thickness, the disc temperature at $R_0$ takes the temperature
of the inner ring: $T(R_0)=T(R_1)>0$~K; see below and see Fig.6).   
    
We divide the disc model into $N$ rings of radius $R_i$ 
($i=1,2,.. N)$. Each ring has a temperature ($T(R_i)$), 
density and effective vertical gravity obtained from 
the disc model
for a given stellar mass ($M_{\rm wd}$), 
inner and outer disc radii ($R_0$ \& $R_{\rm disc}$), 
and mass accretion rate ($\dot{M}$).  
              
We generate a one-dimensional vertical structure for each ring
using  TLUSTY.  
For the disc ring the input is the local mass accretion rate,
mass of the accreting star, the inner radius of the disc, and radius of the 
ring in units of the inner radius of the disc. 

SYNSPEC is then used to create a spectrum for each ring,
and the resulting ring spectra are integrated into a disc spectrum 
using the code DISKSYN, 
which includes the effects of (Keplerian) rotational broadening,
inclination, and limb darkening.
The general procedure we follow to generate disc spectra using TLUSTY 
can be found in \citet{wad98}, and our current disc spectra modeling 
in the UV is given in \citet{god17}. 

\begin{figure} 
\vspace{-3.5cm} 
\includegraphics[width=\columnwidth]{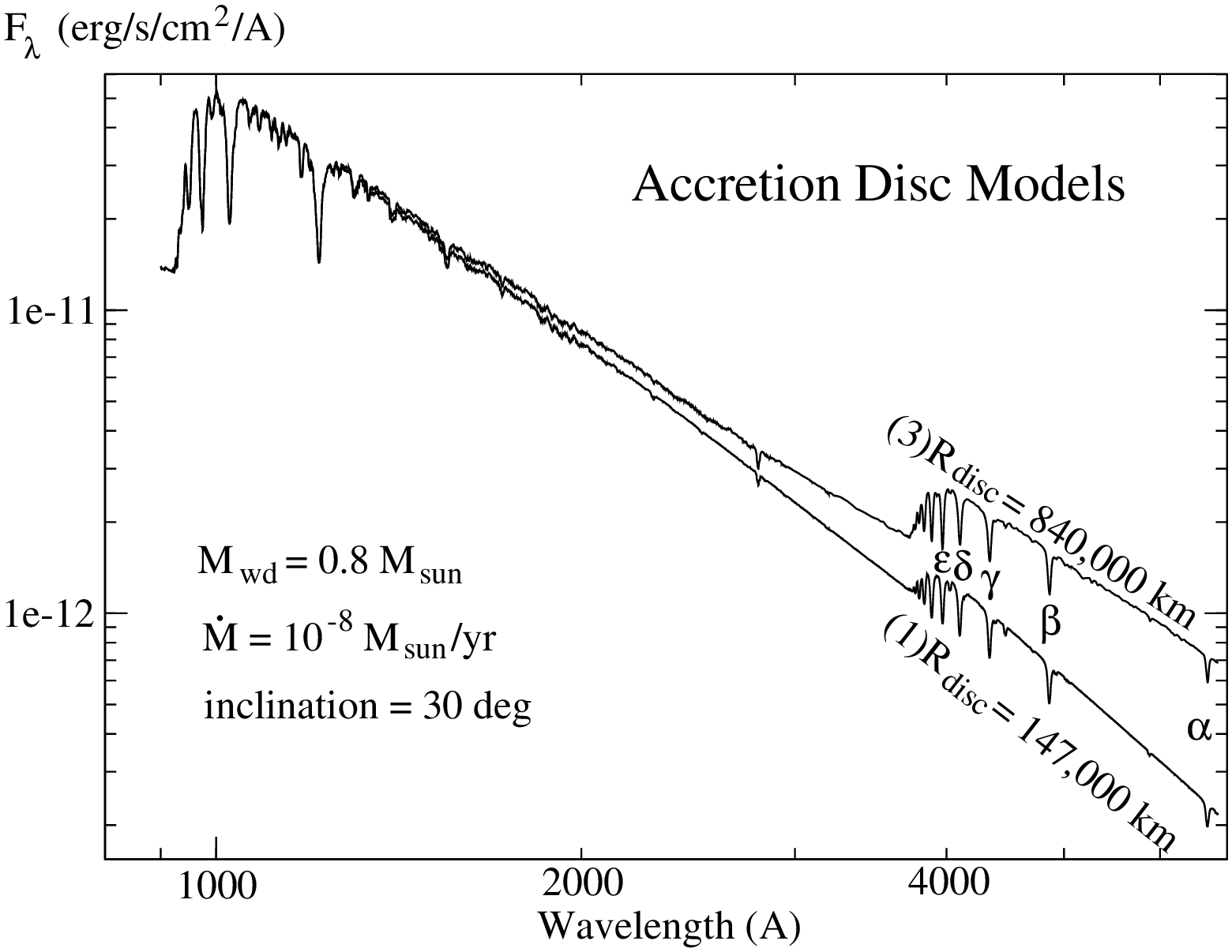} 
\vspace{-1.0cm} 
\caption{
Two accretion disc spectra ($F_{\lambda}$ vs $\lambda$ on a log-log scale) 
generated using TLUSTY/SYNSPEC are displayed. The two discs are identical
(with  a central $0.8M_{\odot}$ white dwarf, a mass accretion rate
$\dot{M}= 10^{-8}M_{\odot}$/yr, and viewed at an inclination $i=30^{\circ}$
for a distance of 100~pc). 
The outer radius in the first disc has been set to 840,000~km,  
and the outer ring reaches a temperature of 3837~K  
(corresponding to model \# 3 in Fig.\ref{disc_temp}).  
In the second disc the outer radius has been set to 147,000~km, and the outer 
ring reaches a temperature of 13,342~K  
(corresponding to model \# 1 in Fig.\ref{disc_temp}).  
The amplitude of the Balmer jump as well as
the slope of the continuum in the longer wavelengths both depend  on
the temperature and, therefore, size of the outer disc.  
\label{Balmer_jump} 
}
\vspace{-3.5cm} 
\includegraphics[width=\columnwidth]{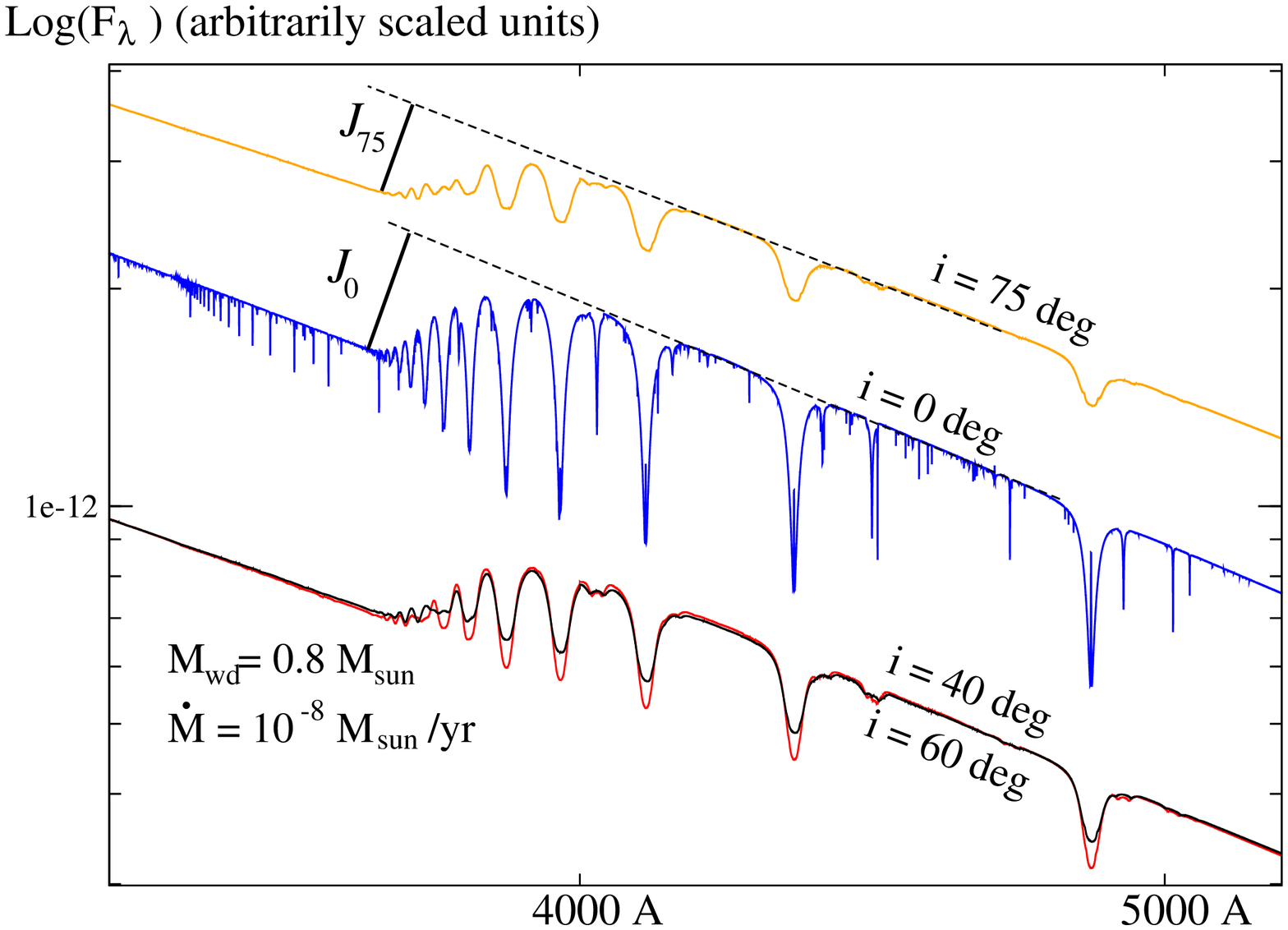}  
\caption{
The Balmer region is shown for four  
inclination angles: $0^{\circ}$ (in blue in the middle),  
$40^{\circ}$ (in red) superposed to $60^{\circ}$ (in black,
bottom), and $75^{\circ}$ (in orange at the top). 
For convenience the spectra have been shifted  
on a log-log scale. 
From $i=0^{\circ}$ to $i=75^{\circ}$ the Balmer jump decreases
by $\sim 25\%$ ($J_{75} \approx 0.75 J_0$), while the absorption
lines decrease by $\sim$80\%. 
For an inclination $i$ varying from $\sim 40^{\circ}$ to $\sim 60^{\circ}$ 
(as considered in the present work),
the {\it relative} depth of the Balmer absorption lines depends strongly 
on the inclination, however, the amplitude of the Balmer jump and the
slope of the continuum are barely affected. 
\label{Balmer_lines} 
}
\vspace{0.5cm} 
\end{figure}

We recently extended our modeling  to include the optical range 
\citep[e.g.][]{dar17,god18}. 
Accretion disc models in the FUV (say 900-2000~\AA ) are usually
limited to an outer radius where the temperature reaches $\sim$10,000~K
(shown in Fig.\ref{disc_temp} with a dotted line),
as lower temperatures contribute very little flux in that spectral
range \citep{wad98}. However, as one considers the optical range up to  
7500~\AA , the accretion disc has to be extended to a larger 
radius to include the colder outer disc (see Fig.\ref{disc_temp}) 
contributing to the optical.  
Here, we incorporate the option to include disc rings as cold as 3,500~K when needed.      
For disc ring temperatures below $\sim$6,000~K (and above $\sim$3,500~K), we 
use Kurucz stellar spectra of appropriate temperature and surface
gravity as mentioned in the previous sub-section.

The important spectral features when modeling the disc in the optical are  
(i) the amplitude of the Balmer jump, 
(ii) the depth of the hydrogen Balmer lines,
and (iii) the slope of the continuum.   
 
We first check the effect of the size of the disc (i.e. $R_{\rm disc}$) 
for an $0.8 M_{\odot}$ WD
\citep[nearly the average CV WD mass][]{zor11} accreting at 
a rate $\dot{M}=10^{-8}M_{\odot}$/yr (for a DN in outburst).  
The temperature profile for this disc model is shown in Fig.\ref{disc_temp}
(upper graph). 
We consider two outer disc radii: 
$R_{\rm disc}=147,000$~km (\# 1 in Fig.\ref{disc_temp}) with a temperature
$T=13,342$~K; and 
$R_{\rm disc}=840,000$~km (\# 3 in Fig.\ref{disc_temp}) with a temperature  
$T=3837$~K. 
We find (see Fig.\ref{Balmer_jump}) 
that the amplitude of the Balmer jump decreases significantly 
as the radius of the disc is decreased from 840,000~km to 
147,000~km, 
corresponding to an increase of almost 10,000~K 
of the outer disc temperature. 
This is due to the fact that model (1) has a temperature well 
above 10,000~K everywhere in the disc (Fig.\ref{disc_temp}).
In addition, the slope of the continuum
in the optical (and somewhat in the near UV) becomes steeper
as the outer radius of the disc decreases, as already noted
by \citet{has83}, but little change occurs in the FUV.  
However, the depth of the Balmer absorption lines remains nearly the same. 
We note that for a secondary mass $M_2=0.3 M_{\odot}$, and assuming 
an outer disc radius of about $a/3$ (33\% the binary separation),
the discs shown in Fig.\ref{Balmer_jump} would belong to a binary
with an orbital period of 19~hr (larger disc) and 1.5~hr (smaller disc).  

Next, we check how the inclination affects the disc spectra.    
We find (see Fig.\ref{Balmer_lines})
that the Balmer jump is not affected by the inclination
as much as the depth of the hydrogen Balmer lines, which 
strongly decrease with increasing inclination. 
In the disc modeling there are three main factors     
affecting the spectrum as the inclination (i) increases: 
(a) the geometric foreshortening of the flat disc 
(reducing the continuum flux level by a factor $\mu = cos(i)$);  
(b) the dependence of specific intensity ($I_{\nu} (\mu)$) on 
the inclination (due to scattering) ;   
and (c) the Keplerian velocity broadening 
\citep[see][]{lad89b,wad98,hub17b}.
The depth of the absorption lines are affected mainly by (c) 
and also to some extent by (b); while the {\it relative} size of the
Balmer jump is affected only by (b), which is stronger at higher
inclinations. 

Consequently, as the mass accretion rate is found mainly by fitting
the FUV for a given white dwarf mass and inclination, the inclination 
(if unknown) and the size of the disc can be found when extending the
fit into the optical.

\subsection{Fitting Technique.} 

In our past research we have used three methods to find the model 
that best fits the data. 

(i) The chi square ($\chi^2$) method, which is meaningful when a 
$\chi^2$ value is found to be significantly smaller (in a statistical sense)
for one model compared to the other models. Because this method is 
quantitative it is often the {\it preferred} one.  

(ii) The best-fit model can also be found by visual inspection of the
fit \citep[see e.g.][]{lin07,lin08,lin09,lin10}. 
This qualitative method is used with common sense and can be as 
effective as (i):
it can lead to the same results within 
the errors and/or uncertainties.
Namely, the models that visually appear to deviate from the observed
spectrum give significantly higher $\chi^2$ values 

(iii) Fitting all the parameters simultaneously. When the systems parameters
(inclination, distance, reddening, WD mass and radius, ..) are known
(with a given accuracy), then the best fit is simply the model which
scales to the known distance (for e.g. for a given WD temperature for 
a DN in quiescence, or for a given mass accretion rate for a disc dominated
system).  

More often than one would like, there can be a discrepancy between the
best fit obtained from (i) or (ii) 
and that derived from (iii). For example, 
the FUV analysis of VW Hyi after a superoutburst \citep{lon09} revealed 
that the scaled radius of the WD is smaller than expected (and even more
so with new Gaia distance of 54~pc). 
And more importantly, the fact that (distance-scaled) disc model fits 
to disc-dominated CVs are too blue compared to UV spectra 
(see Introduction) is itself
a sign that the scaled model fits do not provide the least chi square
fits, or alternatively: the least chi square model fits give the wrong
distance. It is important to also note that for UZ Ser, 
method (i) yielded an inclination $i=18^{\circ}$ in \citet{ham07}
and $i=75^{\circ}$ in \citet{lak01}.  

For best results, either technique (i) or (ii) has to be 
combined with technique (iii). 
 
In the present work we pay particular
attention to the continuua: we choose the model fit exhibiting a
continuum that overlaps the observed the continuum (i.e. we do not 
inspect the absorption lines fit unless we explicitly mention that 
as in sec.5.2.2.), 
especially in those portions of the spectrum that have a higher S/N 
(such as the IUE SWP when compared to the IUE LWR). We also ignore
the emission lines and the contaminated Lyman $\alpha$ region.  
Consequently, in the present work we choose to 
use method (ii) and (iii) to derive the goodness
of fit, namely we use a {\it qualitative} method rather than 
a {\it quantitative} one.   

Using method (iii), the uncertainties of the system parameters
translate into uncertainties in the derived mass accretion rates
and WD temperature. These uncertainties, which we add using 
the $\pm$ sign, reflect the approximate range over which 
good fits are found over the range of values of the parameters
(these uncertainties are not Gaussian distributed errors).   

\section{Results and Discussion} 

\subsection{UZ Ser} 

\subsubsection{The Sep 1981 peak outburst: the SWP15078 spectrum.} 
We start by considering the highest flux IUE spectrum of UZ Ser 
obtained on Sep 22, 1981. 
\citet{ham07} first modeled the SWP15078 segment with a disc 
from the grid of disc models of \citet{wad98}. 
The  Wade \& Hubeny (WH for short)   
disc models are computed
for a WD mass taking values of 0.35, 0.55, 0.80, 1.03, and 1.21$M_{\odot}$,
and the inclination is fixed to i=18$^{\circ}$, 41$^{\circ}$, 60$^{\circ}$, 
75$^{\circ}$, and 81$^{\circ}$. 
Using the $1.03M_{\odot}$ WD mass disc models (with a radius of 5180~km), 
\citet{ham07} obtained a mass accretion rate  
$\dot{M}= 2 \times 10^{-8} M_{\odot}$/yr scaling to a distance of 300~pc. 
In their modeling, they didn't include the possible contribution of a WD.
The reddening they adopted, E(B-V)=0.35, is more likely the upper limit, 
and is much larger than the average \citep[E(B-V)=0.24][]{bru94},   
and the inclination $i=18^{\circ}$ is too low. 
As a consequence, the mass accretion rate derived by Hamilton et al.
is likely overestimated.  

\begin{figure}
\vspace{-3.0cm} 
\includegraphics[width=\columnwidth]{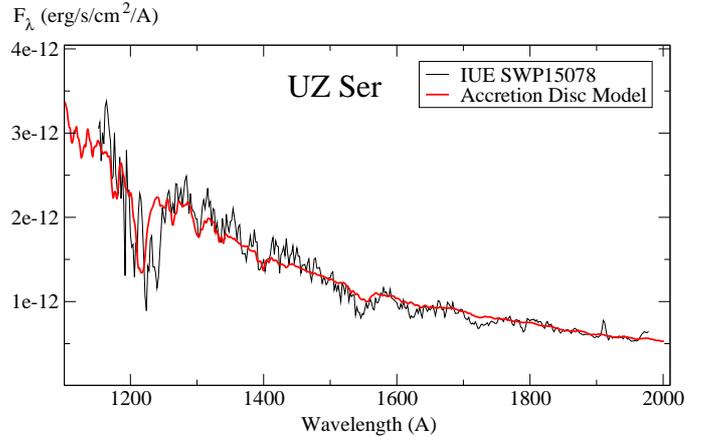} 
\vspace{-1.5cm} 
\caption{
The Sep 1981 outburst spectrum of UZ Ser ({\it IUE} SWP15078, in black) 
is modeled with a Wade \& Hubeny  
accretion disc model (in red). 
The disc model has a $1.03 M_{\odot}$
WD, a mass accretion rate of $\dot{M}=1 \times 10^{-8} M_{\odot}$/yr,  
and an inclination of $41^{\circ}$, giving a distance of 402~pc.  
The inner radius of the disc corresponds to a WD radius of 5,000~km,
and the outer disc radius is 230,000~km, or about 0.23a. 
\label{u_d1.0} 
}
\end{figure}

We first carry out a modeling of the SWP15078 spectrum 
assuming E(B-V)=0.24, and also using a WH disc model, for 
comparison and consistency with 
Hamilton et al. 
We find that   a $1.03 M_{\odot}$ WD accreting at a rate 
of $1 \times 10^{-8} M_{\odot}$/yr, 
with $i=41^{\circ}$ provides a good fit to the spectrum,
giving a distance of 402~pc. 
We  display this model in Fig.\ref{u_d1.0}.  

It is important to note that since the distance to UZ Ser is 
only an estimate, $d \sim 300-400$ pc, the uncertainty in the distance
translates into an uncertainty in the mass accretion 
rate $\dot{M}$ during outburst and an uncertainty in the WD effective
surface temperature $T_{\rm wd}$ at quiescence.  
The same is true for the uncertainties in the WD mass 
($M_{\rm wd} =0.9 \pm 0.1 M_{\odot}$) and the 
extinction value ($E(B-V) \approx 0.25 \pm 0.05$).
The uncertainty in the inclination, $i=50^{\circ} \pm 10^{\circ}$, 
translates only in an uncertainty in $\dot{M}$.  


We ran additional models within the range of the uncertainties of
the parameters. 
Namely, we ran WH models with 
$i=41^{\circ}$, $i=60^{\circ}$ for a distance $d \approx $300, 350, \& 400~pc.
Since the resulting mass accretion rate varies as a function of
the WD mass, inclination and distance assumed in the model,
we write    
$$ 
\dot{M} \equiv \dot{M} (M_{\rm wd},i,d). 
$$ 
For the above range of parameters it becomes
$$
\hspace{-4.cm} 
\dot{M}(1 M_{\odot},
50^{\circ} \pm 10^{\circ},d) \approx 
$$
\vspace{-0.5cm} 
\begin{equation} 
~~~~~~~~~~~~~~~~~~(1 \pm 0.25) \left( \frac{d}{350~{\rm pc}} \right)^2  
 10^{-8} M_{\odot}{\rm yr}^{-1},  
\end{equation} 
where we assume a mean distance of $\sim 350$~pc, 
$1.03 M_{\odot} \approx 1 M_{\odot}$, and $41^{\circ} \approx 40^{\circ}$.
A change of $\pm 10^{\circ}$ (uncertainty) 
in the inclination produces a change of 
about $\pm 1/4$ ($\pm 0.25$) in the mass accretion rate.

Due to the steep slope of the observed SWP IUE spectrum,
the best fits are obtained for the models in eq.(1) with a combination
of parameters ($i$,$d$) giving  the largest  
mass accretion rate $\dot{M} \approx 10^{-8}M_{\odot}$/yr and higher. 
Namely 
$i=40^{\circ}$ with $d=400$~pc,  
$i=50^{\circ}$ with $d \geq 350$~pc,  
and $i=60^{\circ}$ with $d \geq 300$~pc,  
all these models 
give a goodness of fit similar to the model fit displayed in Fig.\ref{u_d1.0}. 

The radius of the WH disc model is $2.3 \times 10^{10}$cm,
corresponding to 0.23$a$ ($a$ is the binary separation)
rather than $a/3$.  
We also tried a disc model we generated from scratch 
with the same parameters and with an outer radius $\sim a/3$, 
but the continuum slope of such a model was too shallow.  
The need for a smaller radius arises from the steep slope of the
continuum, which can only be fitted if the colder (outer) region
of the disc is cut off.

Since the WD mass of UZ Ser is $M_{\rm wd} =0.9 \pm 0.1 M_{\odot}$,  
we also checked a WH disc model with a $0.8 M_{\odot}$ WD 
accreting at $\sim 10^{-8}M_{\odot}$/yr and found that it
does not provide a continuum 
slope steep enough to fit the IUE spectrum. 
As mentioned in section 4.3, all the WH disc models have an
outer disc radius cut off at about $T \sim 10,000$~K 
(see Fig.\ref{disc_temp}).      
To deepen our analysis, we decide to extend the present modeling 
to the longer wavelength (NUV) by including the LWR segment of
the IUE spectrum.

\subsubsection{The Sep 1981 peak outburst: SWP15078 + LWR11605.} 
\begin{figure}
\vspace{-4.0cm} 
\includegraphics[width=\columnwidth]{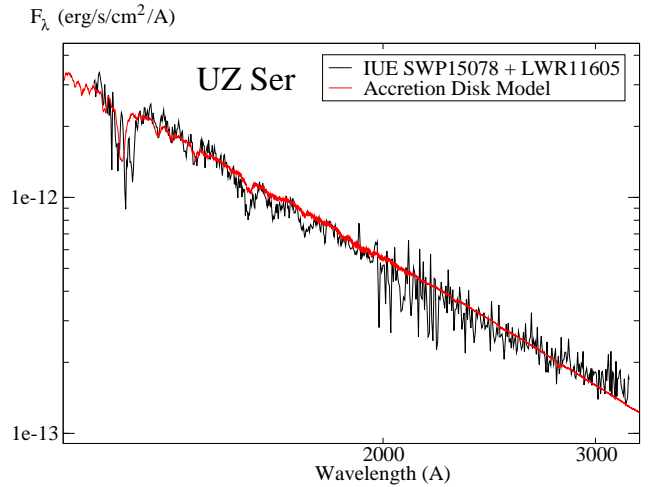}   
\caption{
The {\it IUE} outburst spectrum of UZ Ser (same as in Fig.\ref{u_d1.0})
includes here the longer wavelength (LWR11605) and is fitted with an
accretion disc model. The model has $M_{\rm wd} = 1.0 M_{\odot}$,
$\dot{M}=10^{-8}M_{\odot}$/yr, $i=40^{\circ}$, giving a distance
$d \sim 400~$pc. For clarity the axis are logarithmic. 
Note that in this model the outer radius has been set to 
$0.155a$.  
\label{u_d1.2}   
}
\end{figure}

As the WH disc model spectra do not extend beyond 2000~\AA , 
we generated disc models from scratch 
to check the LW segments of the IUE spectra.  
We find that the combined IUE SWP15078 + LWR11605 (Sep 22, 1981) spectrum 
down to $\sim 3,200$~\AA\ cannot be fitted with the same disc parameters 
(i.e. $M_{\rm wd}=1M_{\odot}$, $\dot{M}=10^{-8}M_{\odot}$/yr, $i=40^{\circ}$,
$R_{\rm disc}=0.23a$),
as the slope of the synthetic spectrum is too shallow 
in the longer wavelengths compared to the LWR IUE spectrum. 
To increase the slope of the disc spectrum 
we have to decrease the outer radius of the disc 
in the model.  
We obtain an outer disc radius of $0.155a$ for a $1.0 M_\odot$ WD
accreting at a rate of $1 \times 10^{-8} M_{\odot}$/yr, 
with $i=40^{\circ}$, giving a distance of $\sim 400$~pc.
This model fit is displayed in Fig.\ref{u_d1.2}.  
For a $0.8M_\odot$ WD disc model, the outer radius of the disc
has to be decreased further to $0.14a$, giving
a distance of 353~pc. 
Everywhere in the disc the temperature is well above 10,000~K,
similar to model \# 1 in Figs.\ref{disc_temp} \& \ref{Balmer_jump}.

Taking the uncertainties of the parameters into account, 
we ran models with $M_{\rm}=0.8M_{\odot}$, $1.0M_{\odot}$, 
$i=40^{\circ}$,  $50^{\circ}$, and $60^{\circ}$, for a distance
$d \approx 300$, 350, and 400~pc. 
Here too the resulting mass accretion rate varies as a function of the
parameters $(d,i)$, and also as a function of the WD mass $M_{\rm wd}$: 
$$
\dot{M} \equiv \dot{M} (M_{\rm wd},i,d).  
$$
Explicitly, we have  
$$  
\dot{M}
(0.9 \pm 0.1 M_{\odot},50^{\circ} \pm 10^{\circ},d) \approx     
$$
\vspace{-0.5cm} 
\begin{equation} 
~~~~~~~~~~~~~~~~~~
(1.1 \mp 0.1 )
(1 \pm 0.25)  \left( \frac{d}{350~{\rm pc}} \right)^2  
 10^{-8} M_{\odot}{\rm yr}^{-1}.  
\end{equation} 
Namely, the change of $\pm 10^{\circ}$ (uncertainty) 
in the inclination produces a change of 
about $\pm 1/4$ ($\pm$0.25) in the mass accretion rate,
and a change of $\pm 0.1 M_{\odot}$ uncertainty in $M_{\rm wd}$ produces
a change of $\mp 10^{-9}M_{\odot}$/yr in $\dot{M}$.
Again, due to the steep slope of the observed IUE spectrum,
the best fit models are for a combination of parameters $(i,d)$ yielding 
a higher mass accretion rate:   
 $\dot{M} \approx 10^{-8}M_{\odot}$/yr and higher.  
Considering also a possible uncertainty in the 
reddening $E(B-V) \approx 0.25 \pm 0.05$ (see end of section 3.1), 
we find an additional uncertainty of +60\% and -40\% in the mass accretion.

The modeling of this first spectrum indicates that the mass accretion
rate near peak outburst is of the order of $10^{-8}M_\odot$/yr, and 
that the observed spectrum is rather blue. 
We also obtained that the outer radius of the disc might
be rather small, however, in order to decide on the size of the outer disc
radius one has to model the hydrogen Balmer region.   
Unfortunately, no simultaneous optical data exist for the Sep 22, 
1981 outburst. 
An additional factor to take into account in the modeling  
is the possible contribution of a hot WD which can increase the steepness 
of the continuum slope to better match the observed (`blue') spectrum.  
The WD is best modeled in quiescence when the contribution from the
disc is negligible, and in the present case, the quiescent spectrum
itself has a rather steep slope. Consequently, we model the Aug 1982  
quiescent spectrum next.

\subsubsection{The Aug 1982 quiescence.}  
The optical quiescent spectrum obtained almost simultaneously with 
the IUE data exhibits many emission lines and a Balmer jump
in emission, pointing to an emitting component which we cannot model
with a WD stellar photosphere or an optically thick disc.
Therefore, we concentrate on 
modeling the quiescent IUE spectrum (SWP17700 and LWR13960) 
obtained on Aug 15, 1982. 
As pointed out earlier, 
the quiescent IUE spectrum also exhibits a rather steep slope, 
qualitatively confirming our suspicion that it is dominated by a 
hot component. 

We first ran accretion disc model fits for 
$M_{\rm wd}= 0.8 - 1.0 M_{\odot}$, 
and a binary inclination $i=40^{\circ}-60^{\circ}$,  
and find that disc models yield a  mass accretion rate in the range 
$1-3 \times 10^{-10}M_\odot$/yr.  
However, these low mass accretion rate disc spectra do not provide enough flux
in the short wavelengths, and are far
too ``red'' compared to the observed spectrum. 

Next, we carried out synthetic stellar spectral fits to the 
quiescent IUE spectrum, fitting mainly the continuum  
since the Ly$\alpha$ region is contaminated with air-glow.  
The spectral fits yield a WD temperature  
$T_{\rm wd} \approx 50,000 - 70,000$~K.
The 50,000~K WD model provided a fit as good as
the 70,000~K WD model, and all the models scaled to the distance
we adopted $d \approx 300-400$~pc.   

In Fig.\ref{u_f_all} we display the 60,000~K WD model 
fit to the IUE quiescent spectrum. 
The model fit is not very sensitive to the effective surface gravity
of the WD, say $Log(g) \sim 8.20-8.60$, as the Lyman region is contaminated. 
At a temperature of 60,000~K, the $0.9M_\odot$ WD has a    
radius of 6,930~km, giving $Log(g)=8.4$, and a distance $d = 360$~pc.   
For a temperature of 50,000~K the distance decreases to 289~pc,
and it increases to 392~pc for T=70,000~K. 
For a $1.0M_\odot$ WD ($R_{\rm wd}=6,230$~km at 60,000~K) the
distance decreases by 10\%, 
and for a $0.8 M_\odot $ WD ($R_{\rm wd}=7,709$~km at 60,000~K) the
distance increases by 11\%. 

Overall, 
the single WD spectral modeling 
can be summarized as follows: 
a temperature 
$T_{\rm wd}=50,000 - 70,000$~K, scaling to a distance 
$d=260-435$~pc, for a white dwarf mass 
$M_{\rm wd}=0.8-1.0 M_\odot$. 
The addition of a low mass accretion rate disc 
with $\dot{M}_{\rm Quies} \sim 3\times 10^{-11}
- 1 \times 10^{-10} M_{\odot}/{\rm yr}$     
to the WD model degrades the WD solution. 
It is, therefore, likely that during
quiescence $\dot{M} < 10^{-10}M_{\odot}$/yr,
and probably even lower: $\dot{M} < 3 \times 10^{-11}M_{\odot}$/yr.
As with most DNe, in quiescence the UV spectrum is likely dominated
by emission from the WD, especially if the WD has an elevated temperature.  

The elevated temperature implied from the quiescent spectrum, 
might be a sign that the WD in UZ Ser is still cooling down from
a nova explosion. Two other Z Cam systems have detected nova 
shells: Z Cam \citep{sha07} and AT Cnc \citep{sha12}.

The IUE SWP17700 quiescent spectrum 
was first modeled by \citet{lak01} assuming
that the reddening was negligible, it was later remodeled in      
\citet{urb06} assuming E(B-V)=0.30. The spectrum was fit with 
a 99,000~K WD (with a mass $M_{\rm wd}=1 M_{\odot}$), and a disc
(contributing only 1\% of the flux) with 
$\dot{M}=10^{-10.5}M_{\odot}$/yr and $i=18^{\circ}$, 
for a distance of 280~pc.  
The hotter WD temperature they obtained  
can be attributed to the larger reddening they used.
\citet{urb06} did not include the LWR segment in their modeling.  

In the optical range, our WD model displays a large discrepancy with
the data (see Fig.\ref{u_f_all}) 
which can be attributed to an additional emitting
component, such as a disc wind \citep{mat15},
especially since the hydrogen Balmer lines and Balmer jump
are all in emission.
Part of the discrepancy, however, could also be due to 
a problem with the calibration of the optical data 
as pointed out in \citet{ver84}.  

As explained in Sec.4.3,
because the spectral coverage in the FUV does not go down the Lyman
limit and also because of the relatively low S/N and air-glow contamination
of the IUE spectra, we do not consider the possibility of a two-temperature
WD model, nor that of a hot boundary layer.

Since the quiescent spectrum was obtained following an outburst
for which a spectrum also exists, we decided, for consistency,
to model that outburst spectrum 
taking into account the hot WD component.   
A decline spectrum was also obtained on Aug 11, 1982, and, therefore,
we model both the Aug 1982 outburst and decline spectra next, 
assuming a WD temperature of 60,000~K and a distance of 
$\sim$360~pc. 


\subsubsection{The Aug 1982 outburst and decline.} 
IUE spectra were obtained at outburst on Aug 08, 1982
(SWP17633 \& LWR13901), and  in decline 
on Aug 11, 1982 (SWP17661 \& LWR13922).   
Simultaneous optical data were digitally extracted only for
the  decline phase and no optical spectrum was available for
digital extraction for the outburst phase.  
Based on the results we obtained for the Aug quiescent spectrum, 
we now include a 60,000~K WD as an addition 
to the disc modeling for the outburst and decline spectra. 

We start by fitting WD+disc models to the decline spectrum.  
We now use disc models we generated from scratch 
with an inclination $i=50^\circ$, 
for a WD mass $M_{\rm wd}= 0.8 M_{\odot}$ \& $M_{\rm wd}=1.0 M_{\odot}$. 

\begin{figure*}
\includegraphics[width=\columnwidth]{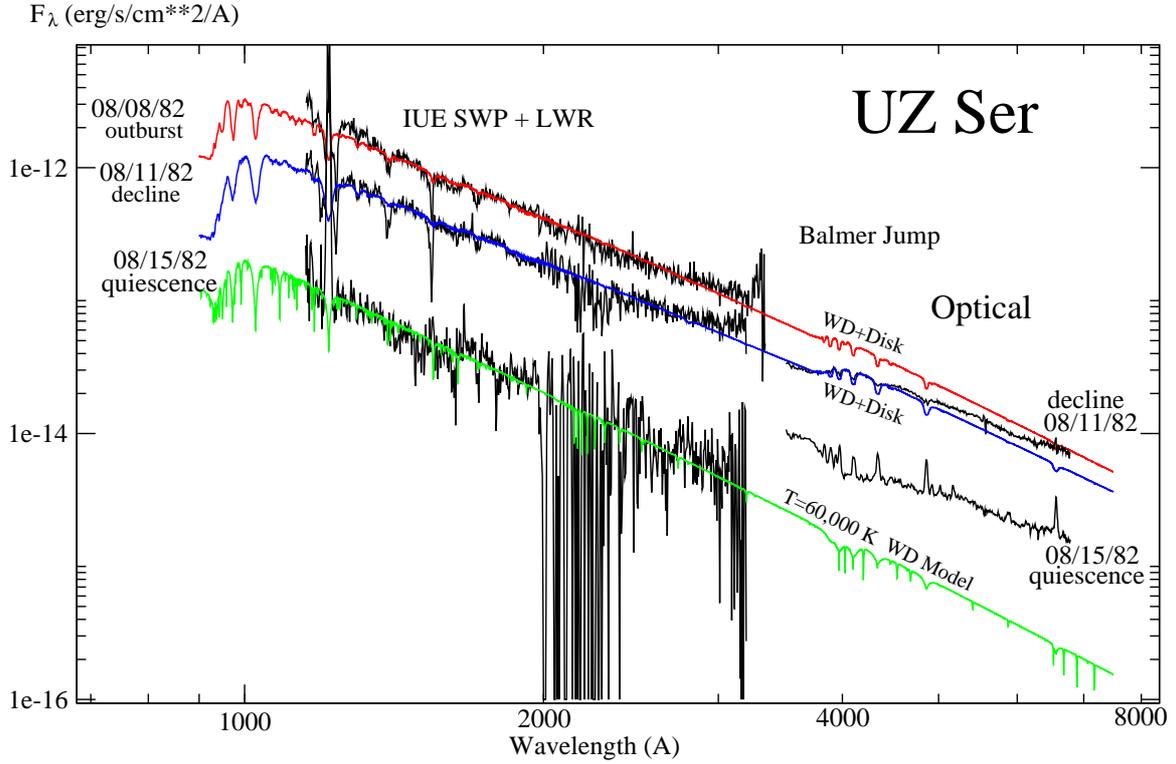}       
\caption{
The observed Aug 1982 UV-Optical spectra of UZ Ser (solid black lines)
are fitted with theoretical model spectra (solid color lines).
Simultaneous IUE and optical spectra were obtained in  
decline (middle) and quiescence (bottom),  
while the outburst spectrum (top) does not have optical data.
In quiescence (Aug 15, 1982) the {\it IUE} UV spectrum 
agrees well with a 60,000~K WD (solid green line), but the optical 
reveals the presence of an emitting component dominated by emission lines.  
Both the outburst and decline spectra are modeled with the combination
a 60,000~K WD 
and an accretion disc 
with a mass accretion rate $\dot{M}\approx 8 \times 10^{-9}M_{\odot}$/yr in outburst,
and $\dot{M}\approx 3 \times 10^{-9}M_{\odot}$/yr in decline. 
In order to fit the Balmer jump and the steep slope of the continuum, 
the outer radius of the disc is set to $R_{\rm disc} \approx 0.1a$ in decline, 
and we assume $R_{\rm disc} \approx 0.2a$ in outburst (see text).  
\label{u_f_all} 
}
\end{figure*}

The outer disc radius is varied  to match the Balmer jump and lines
of the optical spectrum. 
We find that the  mass accretion rate in decline is  
$ 4.1 \times 10^{-9} M_{\odot}/$yr for a $0.8M_\odot$ WD,
with an outer disc radius of 0.1a. 
This model is presented in Fig.\ref{u_f_all}.  
For a   $1.0M_\odot$ WD, the mass accretion rate decreases to 
$ \sim 2.4 \times 10^{-9} M_{\odot}/$yr with an outer disc radius of $0.08a$.  
The very small radius
in these models correspond to model \# 5 in Fig.\ref{disc_temp} 
where the temperature everywhere in the disc is well above 10,000~K.  
The distance fixed for scaling the two models to the data is
$\sim$350~pc.  
Consequently, for a $0.9 \pm 0.1 M_\odot$ WD mass, the mass 
accretion rate in decline is $\dot{M}_{\rm Decl} \approx 3.2 \mp 0.9 \times 
10^{-9}M_\odot$/yr, for a distance of 350~pc.

Next, we fit the Aug 1982 
outburst spectrum with a combined hot WD + disc model. 
For a $0.8M_{\odot}$ WD mass, we find a mass accretion rate  
of $\sim 9.2 \times 10^{-9}M_\odot$/yr
(displayed in Fig.\ref{u_f_all}).  
For a $1.0M_{\odot}$ WD mass, the mass accretion rate decreases 
to $\sim 6.4 \times 10^{-9}M_{\odot}$/yr.   
Here too, the distance was fixed to $\sim$350~pc.
In other words, for a $0.9 \pm 0.1 M_\odot$ WD mass, the mass 
accretion rate in outburst is $\dot{M}_{\rm Outb} \approx  7.8 \mp 1.4 \times 
10^{-9}M_\odot$/yr.
Though there is no optical spectrum for the outburst, an outer
disc radius of $0.14a$ provides a better fitting to the near-UV slope
for an $0.8M_\odot$ accreting WD, and for an $1.0M_\odot$ 
accreting WD the outer disc radius is as large as $0.26a$ 
similar to model \# 1 in Fig.\ref{disc_temp}.  

The addition of the 60,000~K WD barely affects the fit of the  
high mass accretion rate models ($\sim 10^{-8}M_{\odot}$ in outburst),
however, it helps the lower mass accretion rate models 
($\sim 10^{-9}M_{\odot}$ in decline) to better fit the steep
slope of the spectral continuum. 

To summarize, for a WD mass $0.9 \pm 0.1 M_{\odot}$, 
an inclination $i=50^{\circ} \pm 10^{\circ}$,
and a distance of $d$, we find:    
$$               
 \dot{M}_{\rm Outb} \approx  
(8 \mp 1.5) 
(1 \pm 0.25 ) 
\left( \frac{d}{350~pc} \right)^2 
 10^{-9} M_{\odot}{\rm yr}^{-1},  
$$              
\begin{equation} 
\dot{M}_{\rm Decl} \approx  
(3 \mp 1) 
(1 \pm 0.25 ) 
\left( \frac{d}{350~pc} \right)^2 
10^{-9} M_{\odot}/{\rm yr},   
\end{equation} 
$ \dot{M}_{\rm Quies} \lesssim 3  \times 10^{-11} M_{\odot}/{\rm yr},  $ and \\  
$ T_{\rm wd} \approx 60,000\pm 10,000~{\rm K}, $ 
and where we
have rounded up the values to 1 or 2 significant figures.

We find a rather small disc radius during decline by fitting the 
Balmer absorption lines, Balmer jump, and slope of the continuum
in the longer wavelength of the IUE spectrum. 
However, the Balmer H$_\alpha$ and H$_\beta$ lines display broad 
absorption with narrow emission 
and we cannot rule out the possibility that the core of the remaining 
Balmer absorption lines 
might be partially filled in with emission. 
CY Lyr presents a better opportunity to model the outer region of 
the accretion disc.

\clearpage

\begin{figure*}
\includegraphics[width=\columnwidth]{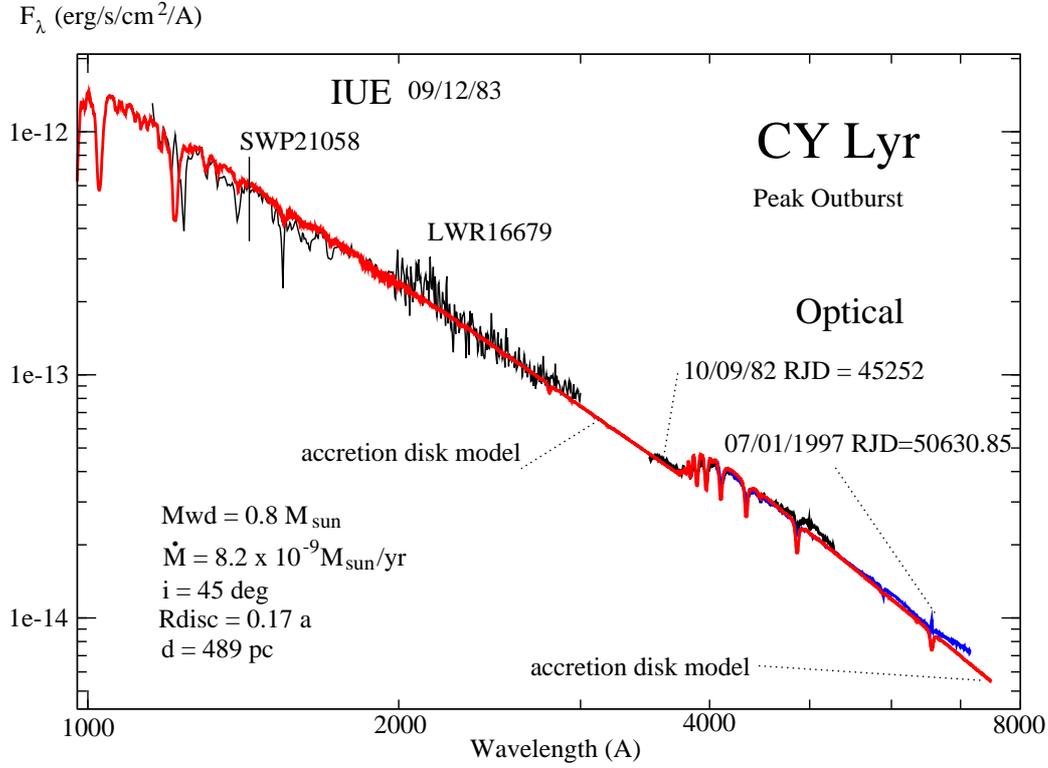} 
\caption{The UV and optical spectra of CY Lyr in outburst
are fit with an accretion disc model. 
The 1983 combined IUE SWP + LWR spectrum (1150-3000~\AA) is in black, the 
1982 optical (z) spectrum 
\citep[$\sim$3,500-5,300~\AA ,][]{szk85} is in black,  
and the 1997 optical (a) spectrum 
\citep[$\sim$4,000-7,200~\AA ,][]{tho98}
is in blue. The disc model (900-7500~\AA) is in red.    
In order to fit the Balmer jump, the outer disc radius has been set to 0.17a. 
For an inclination of $i=45^{\circ}$, one obtains a mass accretion rate of 
$\dot{M}=8.2 \times 10^{-9}M_{\odot}$/yr, and it  
decreases to $\dot{M}=7 \times 10^{-9} M_{\odot}$/yr for an inclination of $i=35^{\circ}$.  
There is a small ``bump'' around 2000~\AA\ which might be mistaken 
for the 2175~\AA\ ISM dust feature 
(the polycyclic aromatic hydrocarbons feature \citep[PAH,][]{li01})  
when over-dereddening the spectrum.
However, the ``bump'' is markedly to the left of the
PAH feature, an indication that it is due to the low S/N 
in that region of the {\it IUE} LWR segment rather than arising from 
over-dereddening the spectrum.  
\label{c_f_all}
}
\end{figure*}

\subsection{CY Lyr} 

\subsubsection{Peak outburst.} 
Although the UV and optical outburst spectra of CY Lyr were not obtained
simultaneously, they all display a consistent matching flux level
(see Fig.\ref{c_all}). 
Consequently, we combine the {\it IUE} UV spectrum together with 
the Oct 9, 1982 \citep{szk85} optical spectrum (z) 
and Jun 30, 1997 \citep{tho98}  optical spectrum (a).
We cut off the {\it IUE} LWR spectrum at 3,000~\AA , since the
edge of the segment is unreliable, and scaled the optical
(a) spectrum to the optical (z) spectrum by multiplying it
by 1.05 (see Table \ref{opt_timing}).   

No UV data exist for the quiescent state of CY Lyr, therefore
we have no way to model the WD. The optical data at quiescence
\citep{tho98} reveals a rather flat continuum dominated by emission
lines, and with a flux level of $\sim$3\% that of the outburst flux.
The quiescent optical spectrum is believed to be from an 
M-dwarf and a low-$\dot{M}$ disc component. 
In the present case we assume that the contribution of the WD
to the UV spectrum in outburst is negligible as well, and we model the combined
UV + optical spectrum with a disc model only.   

The UV region is first fitted with a disc with a given mass accretion rate, 
then the optical region is fitted by varying the size of the outer radius
of the disc. 
As we decrease the outer radius of the disc, the disc temperature
increases, the amplitude of the Balmer Jump in absorption becomes smaller, 
and the slope of the continuum in the optical becomes steeper. 

Since the inclination of the system is unknown, we first assume a
median inclination of $i=45^\circ$, and as stated in section 2, 
we assume a WD with mass of $0.8 M_\odot$, 
and a corresponding radius $R_{\rm wd}=7,000$~km 
(which is also the inner radius of the disc).  
We find that for these parameters, and with a distance of 489 pc,
the UV region matches with a mass accretion rate 
$\dot{M} \approx 8.2 \times 10^{-9}M_\odot$/yr,
while the Balmer jump is matched by truncating the disc at $R_{\rm disc}=0.17a$.   
This model is displayed in Fig.\ref{c_f_all}.  
This disc model is similar to model \# 1 in Fig.\ref{disc_temp},
namely the small outer disc radius is needed to match the Balmer jump
amplitude. 
The UV and optical spectra are compatible with a standard disc
model and exhibit a spectral slope $F_{\lambda} \propto \lambda^{-2.8}$ 
and $F_{\lambda} \propto \lambda^{-3.3}$ respectively.

While we match both the amplitude of the Balmer Jump and the
slope of the continuum, the Balmer absorption lines are not matched: 
the model has deeper lines. However, at peak outburst 
the observed Balmer absorption lines have emission in the their core.   
In a manner similar to UZ Ser, the emission lines might originate
from a disc wind and may also affect the slope and level of
the continuum.  
The optical spectrum obtained during the rise, one day before peak outburst
(spectrum b), exhibits deep Balmer absorption lines without 
peak emission. 
Therefore, we model this spectrum next.

\clearpage 

\subsubsection{Rise to outburst.} 
CY Lyr was caught at the very start of the late June 1997 outburst
and optical spectra were obtained almost every hour covering the
actual rise from quiescence to outburst \citep{tho98}. 
The optical spectrum (b), obtained one day before the outburst spectrum (a),
presents the deepest absorption lines.
It has a continuum flux level of $\sim$66\% the level
reached at outburst, implying that the mass accretion rate 
is half than at outburst.
With a spectral range from $\sim 4,000$~\AA\ to $\sim 7,200$~\AA , 
it does not cover the Balmer Jump. 

We model this spectrum assuming a WD with a mass $M_{\rm wd}=0.8M_{\odot}$,  
a radius $R_{\rm wd}=7,000$~km, and a mass accretion rate $\dot{M}$ of the 
order of a few $10^{-9}M_{\odot}$/yr. 
We have to fit three characteristics of this spectrum: its flux level, 
its slope, and the depth of its absorption lines. 
The flux level and the slope are fitted by varying the mass accretion
rate and the outer radius of the disc until a fit is found.
Then the inclination of the system is varied until the model
fits the depth of the observed absorption lines.  

\begin{figure}
\vspace{-2.0cm} 
\includegraphics[width=\columnwidth]{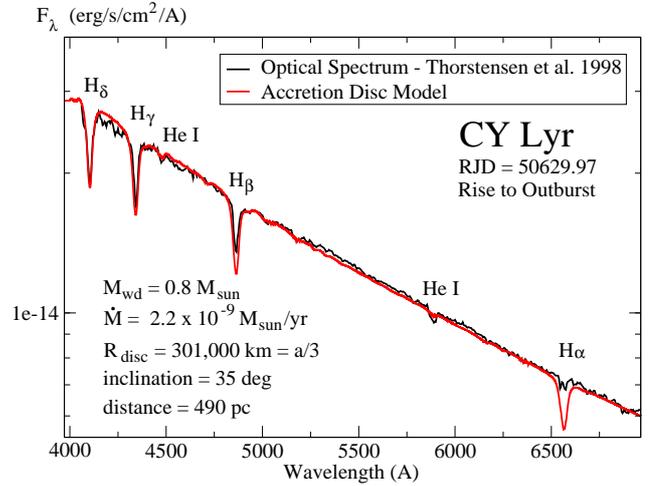}  
\caption{
The 1997 near-outburst optical (b) spectrum 
\citep[$\sim$4,000-7,200~\AA ,][in black]{tho98} is fit 
with an accretion disc model (in red). In order to match
the continuum flux level, the mass accretion rate
has to be of the order of a few $10^{-9}M_{\odot}$/yr. 
The outer radius of the disc is dictated by the slope of the 
continuum and matches the expected theoretical value $a/3$.  
In order to fit the depth of the H$_{\delta}$ and H$_{\gamma}$
absorption lines, the inclination has to be set to $i=35^{\circ}$.
With these parameters, the resulting mass accretion rate is  
$\dot{M}=2.2 \times 10^{-9} M_{\odot}$/yr.   
\label{c_f_i} 
}
\end{figure}

We find a mass accretion rate of  $\approx 2.2 \times 10^{-9}M_{\odot}$/yr 
for a distance of $d=490$~pc, and   
the slope of the optical continuum is best fitted when 
choosing an outer disc radius $R_{\rm disc} = 43 R_{\rm wd} =301,000$~km, or
almost exactly $a/3$. 
This disc model corresponds to model \# 4 
in Fig.\ref{disc_temp} with an outer disc temperature reaching 
as low as 6,000~K.   
The inclination, $i=35^{\circ} \pm 5^{\circ}$, 
was found by  matching the $H_{\gamma}$ absorption lines
(the deepest of all the lines) 
while decreasing the inclination in step of $5^{\circ}$.  
This model fit is displayed in in Fig.\ref{c_f_i}. 
This spectrum with no emission lines 
best represents an optically thick   
standard disc model with deep absorption lines and an outer radius
extending all the way to $\sim a/3$. 

We now remodel the outburst spectrum (Fig.\ref{c_f_all}) with an inclination
$i=35^{\circ}$, and find a slightly lower mass accretion
rate $M_{\rm Outb} \approx 7 \times 10^{-9}M_\odot$/yr. 

For a WD mass of $0.8 \pm 0.1 M_{\odot}$, and inclination
$i=35^{\circ} \pm 5^{\circ}$ we have: 
$$               
 \dot{M}_{\rm Outb}  \approx  
(7.0 \mp 1.3) 
(1.0 \pm 0.1) 
 10^{-9} M_{\odot}{\rm yr}^{-1},  
$$              
\begin{equation} 
 \dot{M}_{\rm rise} \approx  
(2.2 \mp 0.5) 
(1.0 \pm 0.1) 
 10^{-9} M_{\odot}{\rm yr}^{-1}.  
\end{equation}

The other optical spectra on the rise to outburst have some
emission lines contribution. They also have a low
mass accretion rate ($\sim 10^{-9} M_\odot$/yr and lower) 
and, therefore, have a likely contribution from the WD and
secondary star for which we do not have enough data to model.  
In addition, since we do not have any corresponding UV spectra,
we do not attempt to model the remaining optical spectra obtained
during rise to outburst. 

\clearpage 

\section{Summary and Conclusions}

We carried out a combined UV + optical spectral analysis of 
two DNe: UZ Ser in outburst, decline, and quiescence, 
and CY Lyr on the rise to outburst and in outburst. 
These two CVs are among the very few that actually 
seem to be in good agreement with the standard disc model
(with a UV SED $F_{\lambda} \propto \lambda^{-3.0}$ for 
UZ Ser and $\propto \lambda^{-2.5}$, \citep{god17}).

The archival data of these two systems
summarize well the observational properties of DNe. 
In quiescence, while the UV is mostly dominated by emission from
the heated WD, the optical reveals a Balmer jump in 
emission as well as Balmer emission lines which are not due to the WD. 
During the rise to outburst, the spectrum becomes bluer: the disc starts
to contribute more and more to the UV and optical. The Balmer jump and
lines in emission fade and are gradually replaced by a Balmer jump 
and lines in absorption.  The absorption lines are deeper before
the outburst peak. During the outburst peak the optical shows emission
lines in the bottom of the absorption lines.  During decline,
the spectrum displays a Balmer jump
in absorption, but the jump is not very large. The amplitude of the
jump decreased from outburst to decline.

In our modeling, the reduced Balmer jump size (in absorption) 
is due to a small disc radius ($R_{\rm disc} < 0.3a$). 
Using both the IUE and optical data, our results on   
UZ Ser reveal a 60,000~K WD accreting at a rate of  
$\sim 8 \times 10^{-9}M_\odot$/yr in outburst,  
$\sim 3 \times 10^{-9}M_\odot$/yr in decline, 
and more than 100 times lower in quiescence. 
The decline phase with simultaneous UV and optical data 
displays shallow Balmer absorption lines with a small Balmer jump 
in absorption characteristic of a disc extending only to
a radius $\sim 0.1a$. The outburst phase may have a larger
radius of the order of $\sim 0.2a$.  
CY Lyr reveals a similar picture with a mass accretion rate of  
$\sim 8 \times 10^{-9}M_\odot$/yr in outburst,  
and $\sim 2 \times 10^{-9}M_\odot$/yr on the rise to outburst,
but the lack of UV data in quiescence prevents us  
from assessing the WD temperature. 
The modest Balmer jump in absorption during outburst gives
an outer disc radius of $\sim 0.17a$.  
However, the optical data during rise to outburst 
(one day before the outburst peak) 
displaying the deepest absorption lines, gives a disc radius of 
$0.3a$. 

Overall, the results of our spectral analysis show that during 
a dwarf nova cycle the UV-optical spectrum agrees well with 
the standard disc model {\it just before the peak of the outburst}.  
If the depth of the Balmer jump and lines in absorption in UZ Ser is
solely due to the outer disc radius at a fixed inclination, as in our modeling, 
then, during outburst and decline
the disc has a radius smaller than expected ($\sim 0.17a$),  
and one day before the peak outburst, towards the end of the rise, the disc
has a radius of $\sim 0.3a$. 
This consists in a $\sim$43\% change in the disc size relative to is
maximum size.   
This is in {\it qualitative} agreement with the disc instability model (DIM) 
for dwarf nova outbursts \citep{ham98}, 
but quantitatively it over-estimates the relative change.   
Simulations of the DIM 
when taking into account mass-transfer fluctuation due to irradiation
of the secondary \citep{ham00}, or additional dissipation heating
\citep[][stream impact, tidal-torque - see \citet{las01} for a review]{bua01},
exhibit a change in radius size of up to only $\sim 25$\%.

We note that during outburst, the disc models
had a temperature ($T\sim 7000$~K and up) everywhere  
consistent with the upper branch 
of the S-curve of the DIM; during the late rise the disc
model had a minimum temperature reaching 6,000~K; and 
there was no need for models with an outer disc temperature
below that ($T \sim 3500-6000$~K, such as model \# 3 in 
Fig.\ref{disc_temp}) 
as found in the lower branch of the S-curve of the DIM 
\citep{has85}.

The observed reduced Balmer jump size and depth of the Balmer absorption 
lines could also be due, in part, to  a disc wind causing core emission
\citep[as in nova-likes][]{mat15}. 
In quiescence, Balmer emission lines are 
dominating the spectrum, as the system rises to outburst,
the slope of the increasing spectrum steepens and core emission is seen 
at the bottom of the absorption lines. During late rise the absorption
lines are deeper and core emission seems minimal or null.
As the system
reaches its outburst peak the depth of the absorption lines is again 
reduced by core emission. 
We do not rule out the existence of such a wind which
also reduces the slope of the UV and optical continua.
Both the wind and radius
of the disc have to be taken into account for a more
realistic modeling of disc-dominated CV systems - including nova-likes. 

Only at the very end of the rise to outburst does a DN system have  
a spectrum consistent with an actual steady-state standard accretion disc.    
We emphasize that optical data obtained at the end of the 
rise to outburst, especially when combined with available UV data, are extremely valuable
for modelling dwarf novae, as they allow a determination of the inclination of the system, 
the mass accretion rate and the outer disc radius. 

\section*{Acknowledgements}

We wish to thank an anonymous referee for her/his pertinent report
and constructive criticism, which helped improve the manuscript. 
This work is supported by the National Aeronautics and Space 
Administration (NASA) under grant number NNX17AF36G
issued through the Office of Astrophysics Data Analysis Program (ADAP) 
to Villanova University. 
PS acknowledges support from NSF AST-1514737. 
We have made use of online data
from the AAVSO, and we are thankful to the AAVSO staff and members worldwide.



\bsp	
\label{lastpage}
\end{document}